\documentclass[aps,prd,reprint,superscriptaddress,nofootinbib,amsfonts,amssymb,amsmath]{revtex4-1}

\usepackage{blindtext}
\usepackage{lipsum}
\usepackage[utf8x]{inputenc}
\usepackage[dvipdf,dvips,dvipdfmx]{graphicx}
\usepackage{float}
\usepackage{wrapfig}
\usepackage{bm}
\usepackage{color}
\usepackage{comment}
\usepackage{xcolor}
\usepackage[normalem]{ulem}
\usepackage{hyperref}
\hypersetup{
colorlinks=true,
citecolor=blue,
citebordercolor=red,
linktoc=all,
linkcolor=blue,
urlcolor=blue
}

\allowdisplaybreaks

\catcode`\@=11
\def\lsim{\mathrel{\mathpalette\@versim<}}
\def\gsim{\mathrel{\mathpalette\@versim>}}
\def\@versim#1#2{\vcenter{\offinterlineskip
\ialign{$\m@th#1\hfil##\hfil$\crcr#2\crcr\sim\crcr } }}
\catcode`\@=12
\newcommand{\Slash}[1]{{\ooalign{\hfil/\hfil\crcr$#1$}}} 
\newcommand{\bvec}[1]{\mbox{\boldmath $#1$}}

\newcommand{\p}{\partial}

\newcommand{\R}{\text{R}}

\newcommand{\mathL}{\mathcal{L}}

\newcommand{\al}[1]{\begin{align}#1\end{align}}
\newcommand{\bp}{\begin{pmatrix}}
\newcommand{\ep}{\end{pmatrix}}
\newcommand{\nn}{\nonumber\\}

\newcommand{\df}{\text{d}}

\newcommand{\bs}[1]{\boldsymbol}

\newcommand{\pmat}[1]{\begin{pmatrix}#1\end{pmatrix}}

\newcommand{\fn}[1]{\!\left(#1\right)}

\graphicspath{{./figs/}}
\newbox{\ORCIDicon}
\sbox{\ORCIDicon}{\large
                  \includegraphics[width=0.8em]{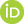}}

\begin{document}
\begin{flushright}
KUNS-2786/KYUSHU-HET-208
\end{flushright}
\vspace{-1cm}
\title{Scalegenesis and fermionic dark matters in the flatland scenario}

\author{Yu \surname{Hamada}\,\href{https://orcid.org/0000-0002-0227-5919}
                               {\usebox{\ORCIDicon}}
	}
\email{yu.hamada@gauge.scphys.kyoto-u.ac.jp}
\affiliation{Department of Physics, Kyoto University, Kyoto, 606-8502, Japan}

\author{Koji \surname{Tsumura}\,\href{https://orcid.org/0000-0003-3765-2750}
                               {\usebox{\ORCIDicon}}
	}
\email{tsumura.koji@phys.kyushu-u.ac.jp}
\affiliation{Department of Physics, Kyushu University, 744 Motooka, Nishi-ku, Fukuoka, 819-0395, Japan}

\author{Masatoshi \surname{Yamada}\,\href{https://orcid.org/0000-0002-1013-8631}
                               {\usebox{\ORCIDicon}}
	}
\email{m.yamada@thphys.uni-heidelberg.de}
\affiliation{Institut f\"ur Theoretische Physik, Universit\"at Heidelberg, Philosophenweg 16, 69120 Heidelberg, Germany}

\begin{abstract}
We propose an extension of the standard model with Majorana-type fermionic dark matters based on the flatland scenario where all scalar coupling constants, including scalar mass terms, vanish at the Planck scale, i.e. the scalar potential is flat above the Planck scale.
This scenario could be compatible with the asymptotic safety paradigm for quantum gravity.
We search the parameter space so that the model reproduces the observed values such as the Higgs mass, the electroweak vacuum and the relic abundance of dark matter.
We also investigate the spin-independent elastic cross section for the Majorana fermions and a nucleon. 
It is shown that the Majorana fermions as dark matter candidates could be tested by dark matter direct detection experiments such as XENON, LUX and PandaX-II.
We demonstrate that within the minimal setup compatible with the flatland scenario at the Planck scale or asymptotically safe quantum gravity, the extended model could have a strong predictability.
\end{abstract}
\maketitle

\section{Introduction}
With the discovery of the Higgs boson~\cite{Aad:2012tfa,Chatrchyan:2012xdj} the standard model (SM) was complete. 
This brings us to the next stage in elementary particle physics.
One of obvious issues is the lack of a dark matter candidate in the SM.
At the present stage a little fact is known about features of the dark matter as an elementary particle.
In particular, it is not clarified even that the dark matter is either fermionic or bosonic, so that a numerous possibility of dark matter candidates can be considered.
Besides, the nature of the Higgs sector is still unclear although all coupling constants in the SM are determined.
Due to this situation, a scenario, where dynamics of dark matter is related to that of the Higgs field, has been suggested.

Let us here discuss what one expects from the discovered Higgs boson mass.
The observed Higgs boson mass 125\,GeV indicates that the perturbative renormalization group (RG) flow of the Higgs quartic coupling constant with the top quark mass $M_t\simeq 171$\,GeV reaches to zero around the Planck scale $M_\text{pl}$ within the standard model (SM)~\cite{Holthausen:2011aa,Degrassi:2012ry}.
This fact might indicate a compelling evidence for dynamics of particles from a high energy theory including quantum gravity if one assumes that the SM is valid up to $M_\text{pl}$ or effects of new physics do not drastically change dynamics of SM particles.
This fact motivates us to consider the flatland scenario~\cite{Iso:2012jn,Chun:2013soa,Hashimoto:2013hta,Hashimoto:2014ela,Haba:2017wwn} as one of scenarios for an extension of the SM.

The flatland scenario imposes that all couplings for scalar fields, involving scalar masses and the Higgs portal coupling, vanish at the Planck scale, namely the scalar potential becomes flat above $M_\text{pl}$.
Such a scenario might be highly controversial from the viewpoint of low energy physics, whereas this could be a natural condition from the asymptotic safety scenario of quantum gravity~\cite{Hawking:1979ig,Reuter:1996cp,Souma:1999at}.
The existence of a non-trivial ultraviolet (UV) fixed point for gravitational couplings realizes asymptotically safe gravity as a non-perturbatively renormalizable quantum gravity.
Above the Planck scale, RG scalings of couplings are non-trivially modified from the canonical ones due to the anomalous dimension induced by graviton fluctuations.
For a certain fixed point value, the scalar masses and the scalar couplings are suppressed and then these couplings become irrelevant parameters~\cite{Wetterich:2016uxm,Eichhorn:2017als}.
This fact enforces scalar interactions so as to be switched off until the Planck scale.

In this work, we propose a U(1)$_X$ extension of the SM compatible with both the flatland scenario and the existence of dark matter candidates.
We add right- and left-handed Majorana fermions, a SM-singlet scalar field which couple to a U(1)$_X$ gauge field.
The Majorana fermions have Yukawa coupling with the singlet scalar field.
These Majorana fermions are stable and thus can become dark matter candidates.
The ratio between the U(1)$_X$ gauge coupling and the Yukawa couplings is a key quantity for the generation of an expectation value of the singlet scalar field (or a finite scale), based on the Coleman-Weinberg mechanism~\cite{Coleman:1973jx}.
As a consequence, the U(1)$_X$ symmetry is broken and then the corresponding gauge boson becomes massive, while the Majorana fermions acquire finite masses via the Yukawa couplings.
In general, it is allowed to write the so-called kinetic mixing term between the U(1)$_Y$ gauge field in the SM and the U(1)$_X$ gauge field.
Such a kinetic mixing plays a crucial role of the inducement of a negative Higgs portal coupling between the Higgs doublet field in the SM and the singlet scalar field.
Thanks to this, the breaking of the U(1)$_X$ symmetry triggers the electroweak symmetry breaking.

The dark matter relic abundance in this model corresponds to that of the Majorana fermions.
This constraint can fix, for instance, the value of the ratio between the Yukawa couplings.
At this point, there is only one free parameter, e.g. the U(1)$_X$ gauge coupling, in this model.
This free parameter could be determined by the direct detection of the weakly interacting massive particle (WIMP).
Hence, this model is testable in near futures.

This paper is organized as follows:
In Section~\ref{Section: Setup}, we briefly explain the basic idea of the asymptotic safety scenario for quantum gravity and its implications for the matter dynamics.
We introduce the model compatible with the flatland condition and summarize both theoretical (from asymptotically safe gravity) and experimental constraints for this model.
We explain the mechanism of the symmetry breaking in the flatland scenario in Section~\ref{sect: Realization of scalegenesis} and show that the electroweak scale is generated within this model by taking a benchmark point.
In Section~\ref{Dark matter analysis} we investigate the relic density of the Majorana fermions as dark matter candidates.
The spin-independent cross section between the Majorana fermions and a nucleon is shown with the current upper bound from the WIMP direct detection experiment.
Section~\ref{sect: summary} is devoted to summarize this work.
In Appendix~\ref{beta functions}, we collect the beta functions at the perturbative one-loop level.
The one-loop effective potential in this model is shown in Appendix~\ref{App: One-loop effective potential}.
We show the explicit forms of cross sections for annihilations of the Majorana fermions in Appendix~\ref{App: Cross sections for dark matter annihilation}.

\section{Setup}
\label{Section: Setup}
In this section, we discuss the basic idea of asymptotically safe quantum gravity and briefly summarize its current status.
In particular, we will stress that quantum graviton fluctuations drive scalar dynamics such that it behaves as a free theory above the Planck scale, and then the flatland condition is given as a consequence from decoupling of quantum gravity effects around the Planck scale.
We propose a flatland model involving fermionic dark matter candidates. 
\subsection{Asymptotic safety scenario}
As a UV-complete theory beyond the Planck scale, we assume asymptotically safe quantum field theory involving quantum gravity.
Here, we start with general discussions on the fixed point structure in a theory space and an energy scaling of a coupling constant in RG flow in order to make our criterion for an extension of the SM.

For a theory space spanned by a set of effective operators ${\mathcal O}_i$ whose (dimensionless) coupling constants are denoted by $g_i$, one explores fixed points at which all beta functions $\beta_i(\{g_i\})$ vanish.
One can easily find the Gaussian (or trivial) fixed point $g_{i*}=0$ which can be discussed by perturbative RG.
In addition to such a fixed point, in several quantum systems, non-trivial fixed point $g_{i*}\neq0$ could exist.
One of well known cases is the Wilson-Fisher fixed point~\cite{Wilson:1971dc} in the three dimensional scalar theory which describes a ferromagnetic phase transition.
In this case, however, non-perturbative methods, e.g. $\epsilon$-expansion~\cite{Wilson:1971dc,Wilson:1973jj} and functional RG~\cite{Polchinski:1983gv,Wetterich:1992yh,Berges:2000ew,Pawlowski:2005xe,Gies:2006wv} should be employed to analyze the fixed point structure due to the strongly correlated system.

Once one finds a fixed point, the energy scaling of coupling constants $g_i$ at vicinity of the fixed point can be clarified.
This is characterized by the critical exponent $\theta_i$ such that the dimensionless coupling constant behaves as $g_i(k)\sim k^{-\theta_i}$ in the RG flow, where $k$ is the energy scale.
More specifically, one can obtain critical exponents by evaluating the eigenvalues of the stability matrix $T_{ij}=\p \beta_i/\p g_j|_{g=g_*}$.
Coupling constants with positive critical exponents are relevant parameters and grow up toward low energy regimes.
The subspace spanned by relevant operators, which are called the critical surface, defines a UV complete renormalizable theory.
The relevant coupling constants are free parameters, so that a system with a less number of relevant couplings has a higher predictability.
In contrast, for negative critical exponents, coupling constants are irrelevant parameters which are driven as functions of relevant couplings.
At the Gaussian fixed point, the energy scaling for an operator can be read as the canonical dimension of its coupling constant, while at the non-trivial fixed point, a large anomalous dimension is induced by quantum dynamics, so that the energy scaling of coupling constants highly deviates from the canonical one.
In such a case, one has to evaluate the eigenvalues of the stability matrix in order to obtain the critical exponents.

Let us here turn to the discussion on the basic idea of asymptotically safe quantum gravity.
It is known that quantum gravity based on the Einstein--Hilbert action is perturbatively non-renormalizable due to the requirement of an infinite number of counter terms~\cite{tHooft:1974toh}.
Nevertheless, there is a possibility that quantum gravity could be formulated as a non-perturbatively renormalizable theory which is known as asymptotically safe quantum gravity~\cite{Hawking:1979ig,Reuter:1996cp,Souma:1999at}.
As discussed above, for the asymptotic safety scenario for quantum gravity, the existence of a non-trivial fixed point for gravitational interactions plays a crucial role.
A number of studies utilizing the functional RG method have been performed and have shown evidences for the existence of such a non-trivial fixed point; see reviews~\cite{Niedermaier:2006wt,Niedermaier:2006ns,Codello:2008vh,Reuter:2012id,Percacci:2017fkn,Eichhorn:2017egq,Eichhorn:2018yfc,Reuter:2019byg}.
An important fact is that a finite  number of positive critical exponents ($\theta_i>0$) for gravitational couplings is observed~\cite{Codello:2007bd,Machado:2007ea,Benedetti:2009rx,Benedetti:2009gn,Falls:2013bv,Falls:2014tra,Gies:2016con,Christiansen:2016sjn,Denz:2016qks,Hamada:2017rvn,Falls:2017lst,Falls:2018ylp,deBrito:2018jxt}, and then asymptotically safe quantum gravity could have a predictability to low energy dynamics.

A great interesting question is impacts of quantum gravity effects on matter dynamics. 
Within the asymptotically safe gravity scenario, a large anomalous dimension induced by graviton fluctuations could change drastically scalings of matter couplings above the Planck scale.
Below a transition scale $k_t$ associated with the Planck scale where quantum gravity effects decouple, dynamics of particles may be described by the SM with a simple extension of the SM.
In this view point, the extended system describing the particle dynamics should satisfy the boundary condition given at $k_t$.

\subsection{Criteria for constructing model}
\label{section: Criterion for construction of model}
We discuss constraints from the asymptotically safe quantum gravity scenario on matter dynamics and consider a possible extension of the SM.

As a simple extension of the SM, the inclusion of a singlet scalar field $S$ coupled to the Higgs field can be considered.
We first discuss the conditions for the RG flow of scalar interactions.
The form of the beta function is given by
\al{
\beta_\lambda= \beta_{\lambda,\text{matter}} + f_\lambda \lambda\,.
}
Here, $\lambda$ denotes a scalar coupling such as the quartic and portal coupling constants, and $\beta_{\lambda,\text{matter}}$ includes contributions from matter dynamics, while $ f_\lambda$ represents universal gravity contribution whose form reads \cite{Eichhorn:2017als,Pawlowski:2018ixd}
\al{
f_\lambda\simeq \frac{\tilde G}{8\pi}\left[ \frac{20}{(1-v_0)^2} +\frac{1}{(1-v_0/4)^2}\right]\,,
\label{eq: anomalous dimension for scalar}
}
where $\tilde G=Gk^2$ is the dimensionless Newton coupling constant and $v_0=16\pi G \Lambda_\text{cc} k^2$ is the dimensionless cosmological constant.
Note that the dimensionful Newton coupling constant $G$ is mass-dimension $-2$, while the mass-dimension of the dimensionful cosmological constant $\Lambda_\text{cc}$ is $2$.
Looking for a fixed point at which $\beta_\lambda=0$, one finds the Gaussian fixed point for the scalar coupling, $\lambda_*=0$.
The investigations for such a system indicate the facts that all scalar couplings involving the Higgs portal coupling are irrelevant~\cite{Eichhorn:2017als} since the critical exponent for the scalar coupling is given by $\theta_\lambda\simeq -f_\lambda^*<0$ where we assume that the gravitational couplings have a non-trivial fixed point.
This means that the RG flow for the quartic coupling and the Higgs portal coupling keeps zero until the Planck scale when their fixed point is Gaussian, $\lambda_{H*}=\lambda_{S*}=\lambda_{HS*}=0$.
Thus, we have the boundary conditions at $k_t=M_\text{pl}$ for the scalar interactions such that
\al{
\lambda_H\fn{M_\text{pl}}=\lambda_S\fn{M_\text{pl}}=\lambda_{HS}\fn{M_\text{pl}}=0\,,
\label{boundary condition for scalar interactions}
}
where $\lambda_H$ and $\lambda_S$ are the quartic coupling constants of the Higgs and the additional singlet scalar fields, and $\lambda_{HS}$ is their Higgs portal coupling constant.
These conditions means that the scalar fields behave as free fields above the Planck scale.

Second, we consider the quantum gravity effects on a squared scalar mass parameter.
Its beta function reads
\al{
\beta_{m}=-2\tilde m^2+\beta_{m,\text{matter}}+f_m \tilde m^2\,,
}
where $\tilde m^2=m^2/k^2$ is the dimensionless squared scalar mass parameter.
The first term on the right-hand side is the canonical scaling term which causes the so-called gauge hierarchy problem since it gives a large value of the critical exponent $\theta_m\simeq 2$ for the Gaussian fixed point at which $\beta_{m,\text{matter}}\simeq 0$ and $f_m\simeq0$.
In such a case, the squared scalar mass is relevant, and then its energy scaling is given as a growing up solution below the Planck scale.
For energy regimes above the Planck scale, the critical exponent of the squared scalar mass could change towards a smaller value than canonical one because of the graviton fluctuations ($f_m\neq 0$) in asymptotically safe gravity such that $\theta_m\simeq 2-f_m^*$.
Indeed, the form of $f_m$ is given by the same as Eq.\,\eqref{eq: anomalous dimension for scalar} ~\cite{Pawlowski:2018ixd}.
If a large anomalous dimension $f_m^*>2$ is induced, the critical exponent of the squared scalar mass turns negative and thus the squared scalar mass is not a free parameter.
For the non-trivial fixed point of the squared scalar mass $\tilde m_*^2\neq 0$, the electroweak scale is explained by the resurgence mechanism in a perturbation~\cite{Wetterich:2016uxm}, while for the Gaussian fixed point $\tilde m_*^2= 0$, the squared scalar mass keeps zero within the RG flow until the Planck scale.
So far, the latter case is typically observed by the functional RG analysis~\cite{Eichhorn:2017als,Pawlowski:2018ixd,Wetterich:2019zdo}, so that for the singlet scalar extension of the SM, the squared scalar mass parameters should satisfy
\al{
m_H^2\fn{M_\text{pl}}=m_S^2\fn{M_\text{pl}}=0\,.
\label{boundary condition for scalar mass}
}
This situation corresponds to the so-called classical scale invariance at the Planck scale~\cite{Wetterich:1983bi,Bardeen:1995kv,Aoki:2012xs}.
In this case, the electroweak scale has to be generated by the dimensional transmutation by the Coleman-Weinberg mechanism~\cite{Coleman:1973jx,Meissner:2006zh,Foot:2007iy,Grabowski:2018fjj,Kwapisz:2019wrl} or strong dynamics~\cite{Hur:2011sv,Holthausen:2013ota,Kubo:2014ova,Haba:2015qbz,Kubo:2015cna,Hatanaka:2016rek,Haba:2017wwn,Haba:2017quk,Kubo:2018vdw,Ouyang:2018eub}.
See also \cite{Ishida:2019gri}.
We call the generation of a scale ``scalegenesis" in order to emphasize that a scale invariant theory generates a scale due to its quantum dynamics.

The conditions \eqref{boundary condition for scalar interactions} and \eqref{boundary condition for scalar mass} indicate that the effective scalar potential is flat above the Planck scale.
This is called the flatland scenario~\cite{Chun:2013soa,Hashimoto:2013hta}.
In this scenario, the scalar interactions have to be generated by quantum effects.
With this fact, the following two things should be satisfied: (i) the effective scalar potential is stable; (ii) the electroweak scalegenesis takes place due to the Coleman-Weinberg mechanism in the singlet-scalar sector.
However, they cannot be realized by only the inclusion of a singlet-scalar field to the SM.
In order to obtain the stable scalar potential, the quartic coupling constants have to be positive for large field values.
A Yukawa interaction can play this role since the beta function of the quartic coupling constant includes the term proportional to the Yukawa coupling constant to the fourth power, $\beta_\lambda\supset -y^4$.
The Higgs quartic coupling constant can be realized due to the effect of the top-quark Yukawa coupling constant, while for the singlet-scalar field, an additional fermionic degrees of freedom coupled to the singlet-scalar field is required in order for its positive quartic coupling constant to be generated.
Generally, for the requirement (ii), the quartic coupling in the RG has to be turned to a negative value around energy scales near a vacuum expectation value $\langle S\rangle$.
A new U(1) gauge interaction with the singlet-scalar field could play such a role since the quantum correction $+g^3$ arises from the gauge interaction in the beta function of the quartic coupling constant.
Therefore, for the electroweak scalegenesis in the flatland scenario, the ratio between the Yukawa coupling and the gauge coupling is crucial.
In sect.\,\ref{sect: Condition for scalegenesis in flatland scenario}, the explicit condition for the ratio to realize the  the electroweak scalegenesis is discussed.

Let us here discuss quantum gravity effects on a U(1) gauge coupling and a Yukawa coupling.
For a U(1) gauge coupling, here denoted by $g$, the beta function is given by
\al{
\beta_g=\beta_{g,\text{matter}}-f_g g\,,
}
where the correction from quantum gravity is found \cite{Christiansen:2017cxa} to be
\al{
f_g\simeq \frac{\tilde G}{16\pi}\left[ \frac{8}{1-v_0} -\frac{4}{(1-v_0/4)^2}\right]\,.
}
We should note here that $ -f_g$ takes a negative value for a non-trivial fixed point of the gravitational couplings.
In this case, the matter contribution $\beta_{g,\text{matter}}$ and the gravity effect could balance.
Consequently, one could consider two possibilities as UV complete scenarios~\cite{Harst:2011zx,Eichhorn:2017lry}: One is that in the continuum limit the gauge coupling reaches to a Gaussian fixed point ($g_{*}=0$) at which the gauge coupling is relevant.
In this case, the gauge coupling is a free parameter and behave as an asymptotically free coupling.
Other is the case that a non-trivial fixed point $g_{*}\neq 0$, at which the gauge coupling is irrelevant and asymptotically safe, i.e. the gauge coupling is predictable in the low energy regime.
From these facts, in order for the gauge coupling to be a UV complete coupling, it cannot take a larger value than the non-trivial fixed point value.
For a one-loop level of the beta functions for the gauge coupling, $\beta_{g,\text{matter}}\simeq \beta_{g,\text{1-loop}}g^3$,  the gauge coupling is bounded so that, at the transition scale $k_t=M_\text{pl}$,
\al{
\label{eq: upper bound for gauge coupling}
g\fn{M_\text{pl}} \lsim g_*=\sqrt{\frac{f_g}{\beta_{g,\text{1-loop}}}}\,.
}

In the same manner, the Yukawa couplings could have also an upper bound~\cite{Eichhorn:2016esv,Hamada:2017rvn,deBrito:2019epw}.
On the other hand, the ratio between the Yukawa coupling and the gauge coupling is constrained from the condition for the realization of scalegenesis as will be seen in Section~\ref{sect: Condition for scalegenesis in flatland scenario}.
Together with the bound for the gauge coupling \eqref{eq: upper bound for gauge coupling}, this condition provides both upper and lower bounds for the Yukawa coupling.
Therefore, in this work we do not consider the bound for a Yukawa coupling from the asymptotic safety scenario.

\subsection{Model in flatland}
\label{Sec: Model in flatland}
Following the discussions in the previous subsection, we consider an extension of the SM based on the flatland scenario.
As a possible extension, we propose an extended system with a SM singlet-scalar field and Majorana fermions coupled to an extra U(1)$_X$ gauge field.
This case allows us to write a kinetic mixing~\cite{Holdom:1985ag} between U(1)$_Y$ gauge field in the SM and the additional U(1)$_X$ gauge field such that $B_{\mu\nu}X^{\mu\nu}$, where $B_{\mu\nu}=\p_\mu B_\nu -\p_\nu B_\mu$ and $X_{\mu\nu}=\p_\mu X_\nu -\p_\nu X_\mu$ are the field strengths for the U(1)$_Y$ and U(1)$_X$ gauge fields, respectively.
Then, the kinetic terms for these gauge fields are given by
\al{
{\mathcal L}_\text{gauge}=-\frac{1}{4} B_{\mu\nu}B^{\mu\nu}-\frac{1}{4} X_{\mu\nu}X^{\mu\nu}-\frac{\epsilon}{2} B_{\mu\nu}X^{\mu\nu}\,,
\label{kinetic terms for gauge fields}
}
while the interactions between the gauge fields and a matter field are given through a covariant derivative,
\al{
D_\mu=\p_\mu -iYg_YB_\mu-iY_Xg_X X_\mu\,,
\label{covariant derivative}
}
where the strong (QCD) and weak interactions are omitted.
Here, we canonically normalize the kinetic terms \eqref{kinetic terms for gauge fields} so that
\al{
{\mathcal L}_\text{gauge}=-\frac{1}{4} F'_{\mu\nu}F'{}^{\mu\nu}-\frac{1}{4} G'_{\mu\nu}G'{}^{\mu\nu}\,,
\label{kinetic terms for gauge fields after transformation}
}
where $F'_{\mu\nu}$ and $G'_{\mu\nu}$ are the field strengths for a new gauge-field basis $(B'_\mu,X'_\mu)$ defined by the transformation, 
\al{
\pmat{
B'_\mu\\[1ex]
X'_\mu
}
&=\frac{1}{\sqrt{2}}
\pmat{
\epsilon_- && \epsilon_+ \\[1ex]
-\epsilon_+ && \epsilon_-
}\frac{1}{\sqrt{2}}
\pmat{
\epsilon_-&& -\epsilon_- \\[1ex]
\epsilon_+ && \epsilon_+
}
\pmat{
B_\mu\\[1ex]
X_\mu
}\nn[1ex]
&=
\pmat{
1 && \epsilon\\[1ex]
0 && \sqrt{1-\epsilon^2}
}\pmat{
B_\mu\\[1ex]
X_\mu
},\label{new basis}
}
with $\epsilon_\pm=\sqrt{1\pm \epsilon}$.
The second matrix on the right-hand side in the first line of Eq.\,\eqref{new basis} is employed in order to canonically normalize the kinetic terms for the gauge fields, while the first one corresponds to a rotation with an angle of $\tan^{-1}\fn{\epsilon_+/\epsilon_-}$.
The latter transformation can be performed since the kinetic terms \eqref{kinetic terms for gauge fields after transformation} is invariant under a rotation for a field basis.
We see here that for the new field-basis $(B_\mu',X_\mu')$ defined by Eq.\,\eqref{new basis}, the mixing effect corresponds to a scale transformation for $X_\mu$, whereas $B_\mu$ is transferred to $X_\mu$ through the mixing effect.

In the new field-basis \eqref{new basis}, the covariant derivative \eqref{covariant derivative} becomes
\al{
D_\mu=\p_\mu &-iY(  g_Y B'_\mu +g_{\text{mix}} X'_\mu)
-iY_X   g_X' X'_\mu\,.
\label{covariant derivative in new basis}
}
where we define new gauge coupling constants,
\al{
&g_{\text{mix}}=-\frac{\epsilon}{\sqrt{1-\epsilon^2}}g_Y\,,&
& g_{X}'=\frac{1}{\sqrt{1-\epsilon^2}}g_X\,.&
}
A field for which U(1)$_Y$ hypercharge is assigned interacts with $X_\mu$ even if it has no U(1)$_X$ hypercharge.
Hereafter we work in the field-basis $(B'_\mu,X'_\mu)$ and neglect primes on these fields and the coupling constant $g_X'$.

We here give the Lagrangian for our model,
\al{
\mathL = \mathL_\text{SM}|_{m_H\to0}+\mathL_\text{kin}+\mathL_{\chi}-V\,,
\label{starting action}
}
where $\mathL_\text{SM}$ denotes the Lagrangian for the SM without the Higgs mass term due to the condition \eqref{boundary condition for scalar mass}.
Here, $\mathL_\text{kin}$ involves the kinetic part of new fields, 
\al{
\mathL_\text{kin}=|D_\mu S|^2+\bar\chi_R i\Slash D \chi_R+\bar\chi_L i\Slash D \chi_L
- \frac{1}{4}X^{\mu\nu}X_{\mu\nu}\,,
\label{kinetic part of new fields}
}
where $S$ is a singlet-scalar field.
Majorana fermions $\chi_R$ and $\chi_L$ are introduced in order to avoid the gauge anomaly.
These fields interact with the U(1)$_X$ gauge fields via the covariant derivative given in Eq.\,\eqref{covariant derivative}.
The singlet-scalar field $S$ is coupled to only $X_\mu$ with a hypercharge $Y_X=2$.
For the Majorana fermions $\chi_R$ and $\chi_L$, their hypercharges are assigned as $Y_X=-1$ for the U(1)$_X$ gauge field, but is not for the U(1)$_Y$ gauge field, i.e. $Y=0$.
The assignment of hypercharges $Y$ and $Y_X$ for each field is summarized in Table~\ref{standard model particle content}.
In this setup, the Majorana fermions are stable particles after the breaking of the U(1)$_X$ symmetry into the $Z_2$ symmetry, so that they could be dark matter candidates~\cite{
Radovcic:2014rea,Benic:2014aga,
Kim:2006af,Kanemura:2010sh,Djouadi:2011aa,LopezHonorez:2012kv,deSimone:2014pda,Matsumoto:2014rxa,Alves:2015mua,Escudero:2016gzx,Kearney:2016rng,Alves:2016cqf,Arcadi:2017hfi,Han:2017qkr}.

The Majorana-type Yukawa interactions between $S$ and $\chi_R$ or $\chi_L$ are given by
\al{
\mathL_{\chi}= -y_R S \overline{\chi^c}_R \chi _{R} -y_L S \overline{\chi^c}_L \chi _{L}+\text{h.c.}\,. 
\label{Majorana Yukawa interactions}
}
The U(1)$_X$ symmetry prohibits the Majorana mass terms and the Dirac-type Yukawa interactions, while the Dirac type mass is forbidden by scale symmetry.
After the singlet-scalar field has a finite expectation value $\langle S\rangle$, these terms turn to the Majorana mass terms.
Thus, Eq.\,\eqref{Majorana Yukawa interactions} becomes origins of dark matter masses.

The scalar potential, denoted by $V$ in the Lagrangian \eqref{starting action}, is given by
\al{
V=\lambda_H |H|^4+\lambda_S |S|^4+\lambda_{HS} |H|^2|S|^2\,,
}
where $H$ is the Higgs doublet field. 
With the condition \eqref{boundary condition for scalar mass}, the scalar mass parameters keep zero within their RG flows since the beta functions for the scalar mass parameters are proportional to themselves.
Therefore, the scalar mass terms are not taken into account.
In contrast, the quartic and the Higgs portal interactions are needed in order for the theory to be renormalizable within our Lagrangian \eqref{starting action}.
Nevertheless, the RG equations for their renormalized coupling constants have to satisfy the condition \eqref{boundary condition for scalar interactions}, so that these scalar coupling constants are not treated as free parameters.
Note that a massless scalar theory is renormalizable~\cite{Lowenstein:1975rf}.
\begin{table*}  
\begin{center}
\begin{tabular}{|c||c|c|c|c|} \hline
    field  & SU(3)$_{\rm c}$ & SU(2)$_L$ & U(1)$_Y$  & U(1)$_X$\\ \hline \hline
    $q_L = \pmat{u_L\\ d_L},~\pmat{c_L\\ s_L},~\pmat{t_L\\ b_L} $ & \bf 3 & \bf 2 & $1/6$ & 0 \\ [10pt] \hline
    $u_R = u_\R, c_R, t_R$ & \bf 3 & \bf 1 & $2/3$ & 0 \\  \hline
    $d_R = d_\R, s_R, b_R $ & \bf 3 & \bf 1& $-1/3 $ & 0  \\ \hline
    $\ell _L =\pmat{\nu _{e L} \\ e_L},~\pmat{\nu _{\mu L}\\ \mu_L},~\pmat{\nu_{\tau L}\\ \tau_L}$ & \bf 1 & \bf 2& $-1/2$ & 0 \\  [10pt] \hline
    $e_R= e_R, \mu_R, \tau_R$ & \bf 1 & \bf 2 & $-1$ & 0 \\ \hline
   $H= \pmat{\varphi^+ \\ H^0 } $  & \bf 1 & \bf 2 & 1 & 0 \\ \hline
   $A_\mu^i$ (SU(2)$_L$ gauge field)   & \bf 1 & \bf 3 & 0 & 0 \\ \hline
   $B_\mu$ (U(1)$_Y$ gauge field) & \bf 1 & \bf 1 & 0 & 0 \\ \hline
    $g_\mu^a$ (gluon)& \bf 8 & \bf 1 & 0  & 0 \\ \hline
    $S$ & \bf 1 & \bf 1 & 0  & $2$ \\ \hline
    $\chi_L$, $\chi_R$ & \bf 1 & \bf 1 & 0  & $-1$ \\ \hline
  \end{tabular}
  \caption{Charge assignment for elementary particles}
  \label{standard model particle content}
  \end{center}
\end{table*}

Here, we briefly describe the scalegenesis in our model.  
A scale associated with both the electroweak scale and dark matter mass scale should be generated by radiative corrections, i.e., the Coleman--Weinberg mechanism.
Within our present model \eqref{starting action}, the Coleman--Weinberg mechanism in the singlet-scalar sector first could take place.
We denote the generated vacuum by $v_S=\sqrt{2}\langle S\rangle$.
This generation of a scale triggers the electroweak symmetry breaking through a negative Higgs portal coupling, namely
\al{
v^2_H=-\frac{\lambda_{HS}}{2\lambda_H}v_S^2\,,
}
where $\langle H\rangle=(0,\,v_H/\sqrt{2})^T$.
In order to obtain a finite $v_H$, a negative value of $\lambda_{HS}$ has to be induced by quantum effects.
In the next section~\ref{sect: Realization of scalegenesis}, we will see the occurrence of such a situation.
For $\lambda_{HS}\approx 0$, one obtains the Higgs mass $M_H^2\simeq2\lambda_H v_H^2$.
Since U(1)$_X$ symmetry is spontaneously broken, the $X$ boson obtains a finite mass,
\al{
M_{X}\simeq 2g_X v_S\,.
} 
The masses for the Majorana fermions are given by
\al{
&M_R\simeq \sqrt{2}y_R v_S\,,&
&M_L\simeq \sqrt{2}y_L v_S\,.&
}
The difference between $\chi_R$ and $\chi_L$ in this model is characterized by only the Yukawa coupling constants.
Therefore, one can concentrate on the case $y_R\leq y_L$ with no loss of generality.

Finally, we mention the constrains on parameters involved in the present model.
In addition to the three conditions \eqref{boundary condition for scalar interactions} for the quartic coupling constants and the Higgs portal coupling constant, we have constraints from the observations~\cite{Tanabashi:2018oca}
\al{
&v_H=246\,\text{GeV},\nn[1ex]
&M_H^\text{obs}=125.10 \pm 0.14\,\text{GeV},\nn[1ex]
&M_t^\text{obs}=160^{+5}_{-4}\,\text{GeV}.
\label{observed masses and vacuum}
}
For the top-quark mass, the $\overline{\text{MS}}$ mass is used.
Note that the pole mass is $M_t^\text{pole}=173.1 \pm 0.9\,\text{GeV}$.
As mentioned above and discussed in Section~\ref{Dark matter analysis}, the Majorana fermions could be stable and then become dark matter candidates.
The latest observation~\cite{Aghanim:2018eyx} reports that the dark matter relic density in the cosmological evolution is
\al{
\label{observed relic abundance}
\Omega_\text{DM}^\text{obs}\hat{h}^2=0.1193\pm 0.0014\,,
}
where ``DM" is the abbreviation of dark matter.
The relic density of the Majorana fermions has to satisfy this value.
These constraints reduce from the five free parameters, $g_{X}$, $g_\text{mix}$, $y_t$, $y_L$ and $y_R$, to one parameter.

As discussed in Section~\ref{section: Criterion for construction of model}, there is an upper bound \eqref{eq: upper bound for gauge coupling} for a U(1) gauge coupling.
For the prediction of the observed value of gauge couplings within the SM, the gravitational contribution $f_g$ being of order $10^{-2}$ is required~\cite{Eichhorn:2018whv}. It is shown in Refs.\,\cite{Harst:2011zx,Eichhorn:2017lry,Eichhorn:2017ylw} that gravitational effects actually yield $f_g$ of this order.
We use $f_g=0.02$~\cite{Reichert:2019car}.
Within our model setup \eqref{starting action}, the upper bound becomes 
\al{
g_X\fn{M_\text{pl}}\lsim 1.09\,,
}
where we used the beta function for $g_X$ given in Eq.\,\eqref{App: beta functions for new gauge couplings} and set $g_\text{mix}=0$.
Using the beta function for $g_\text{mix}$ with $g_X=0$  the kinetic mixing is also constrained as
\al{
g_\text{mix}\fn{M_\text{pl}}\lsim 0.68\,.
}

Finally, we comment on phenomenological constraints for the kinetic mixing effect.
The kinetic mixing coupling constant $\epsilon$ in the Lagrangian \eqref{kinetic terms for gauge fields} is constrained for a wide range of the extra gauge boson mass $M_X$~\cite{Jaeckel:2010ni}. 
The $Z'$-boson mass in our system would become typically of order a few TeV.
For such a mass range, the upper bound on $\epsilon$ is given from $Z'$ searches in the LHC experiments by looking at $e^+e^-$ and $\mu^+\mu^-$ channels.
For $M_X=1$\,TeV, we have $\epsilon \lsim 0.1$~\cite{Jaeckel:2012yz}. 
The bound for heavier mass regions is relaxed such that, for instance, $\epsilon \lsim 0.2$ for $M_X=2$\,TeV.

\section{Realization of scalegenesis}
\label{sect: Realization of scalegenesis}
In this section, we study the mechanism of the scalegenesis in our model using the RG.
First, we discuss a general condition to realize the scalegenesis in the flatland, which can be read from coefficients of beta functions at the one-loop level.
We see that our model actually satisfies the condition, and then we discuss how the physical vacuum and masses are defined.

\subsection{Condition for scalegenesis in flatland scenario}
\label{sect: Condition for scalegenesis in flatland scenario}
We start by looking at a general condition to realize the flatland scenario by following the literature~\cite{Hashimoto:2013hta}.
For a system where a fermion and a scalar boson couple to each other and to a gauge boson, the RG equations for the gauge coupling $g$, the Yukawa coupling constant $y$ and the quartic coupling constant $\lambda$ at the one-loop level are given by
\al{
&\mu \frac{\df g}{\df \mu}=\beta_g=\frac{a}{16\pi^2}g^3 +\cdots\,,\\
&\mu \frac{\df y}{\df \mu}=\beta_y=\frac{y}{16\pi^2}\left[ by^2 -cg^2\right]+\cdots,\\
&\mu \frac{\df \lambda}{\df \mu}=\beta_\lambda=\frac{1}{16\pi^2}\left[ -d y^4+fg^4\right]+\cdots,
}
where $\mu$ is a RG scale, $a,\cdots, f$ are coefficients depending on the degrees of freedom of fields and dots stand for irrelevant terms for the discussion below.
As we have discussed in the subsection~\ref{section: Criterion for construction of model}, the ratio between the Yukawa coupling and the gauge coupling, $r=y/g$, is crucial for realization of scalegenesis in the flatland scenario.
Therefore, we rewrite the beta functions in terms of $r=y/g$ such that
\al{
&\mu \frac{\df r}{\df \mu}=\frac{b\,rg^2}{16\pi^2}( r^2-r_c^2)\,,&
&\mu \frac{\df \lambda}{\df \mu}=\frac{dg^4}{16\pi^2}(r_0^4-r^4)\,,&
\label{general RG equations for r and lambda}
}
where
\al{
&r_c=\sqrt{\frac{a+c}{b}}\,,&
&r_0=\left( \frac{f}{d}\right)^{1/4}\,.&
}

Let us here discuss a realization of a stable and finite vacuum generated from the Coleman--Weinberg mechanism in the flatland scenario.
The effective scalar potential should be bounded for a large field value, $\varphi\sim M_\text{pl}$, to realize a stable vacuum.
This requires that $\beta_\lambda<0$ for $\mu=M_\text{pl}$.
The generation of a scale, here denoted by $v$, in the Coleman--Weinberg mechanism is realized by a negative quartic coupling constant at $\mu=v$.
For this, we need $\beta_\lambda>0$ at $\mu=v$.
From the beta function for $\lambda$ in Eq.\,\eqref{general RG equations for r and lambda}, these behavior can be achieved by $r\fn{v}<r_0<r\fn{M_\text{pl}}$.
This condition also requires that $r$ as a function of $\mu$ has to increase with increasing the scale.
Thus, the generation of a scale in the flatland scenario could be realized when the condition $r_c<r\fn{\mu=v}<r_0<r\fn{M_\text{pl}}$ is satisfied.
This condition can be expressed in terms of the coefficients in the beta functions as  
\al{
K=\left( \frac{r_c}{r_0}\right)^2
=\frac{a+c}{b}\sqrt{\frac{d}{f}}<1.
\label{condition for the flatland scenario}
}

In our model \eqref{starting action}, the Coleman--Weinberg mechanism works in the singlet-scalar sector, so that we can identify the coupling constants $g$, $y$, and $\lambda$ with $g_X$, $y_R$ (or $y_L$), and $\lambda_S$, respectively.
In this case, we have, for each beta function, the coefficients as
\al{
&a=\frac{8}{3},&
&b=8,&
&c=6,&
&d=32,&
&f=96,&
}
where we assume that $y_L=y_R$.
See Appendix~\ref{beta functions} for explicit forms of the beta functions.
Inserting these values into Eq.\,\eqref{condition for the flatland scenario}, we obtain $K\simeq 0.625$ and then can see that the condition \eqref{condition for the flatland scenario} is satisfied.
Note that the value of $r\fn{v_S}$ has to be between $r_c\simeq 1.04$ and $r_0\simeq 1.32$.
We also note that for $y_R=0$ one has $b=6$ and $d=16$, and then $K\simeq 0.590$.

\subsection{Scalegenesis in flatland}
We investigate the scalegenesis in the singlet-scalar sector owing to the Coleman-Weinberg mechanism.
To this end, let us start with a discussion of the mechanism for scalegenesis in our model.
As we have seen in Section~\ref{sect: Condition for scalegenesis in flatland scenario}, the ratios $r_{L,R}=y_{L,R}/g_X$ determine the scale of scalar field $S$. This scale is mediated through the Higgs portal coupling $\lambda_{HS}$ which is set to zero at the Planck scale.
In order for the Higgs portal coupling to have a finite and negative value in low energy regimes, one needs a term not proportional to $\lambda_{HS}$ in its beta function.
Indeed, as one can see from Eq.\,\eqref{App: beta functions for scalar couplings} a term $+(g_\text{mix}g_X)^2$ in the beta function plays such a role, so that a finite value of $g_\text{mix}$ is crucial for the inducement of the electroweak scale via the Higgs portal coupling. (See Eq.\,\eqref{generated Higgs mass squared} below.)

\begin{figure}
\includegraphics[width=8cm]{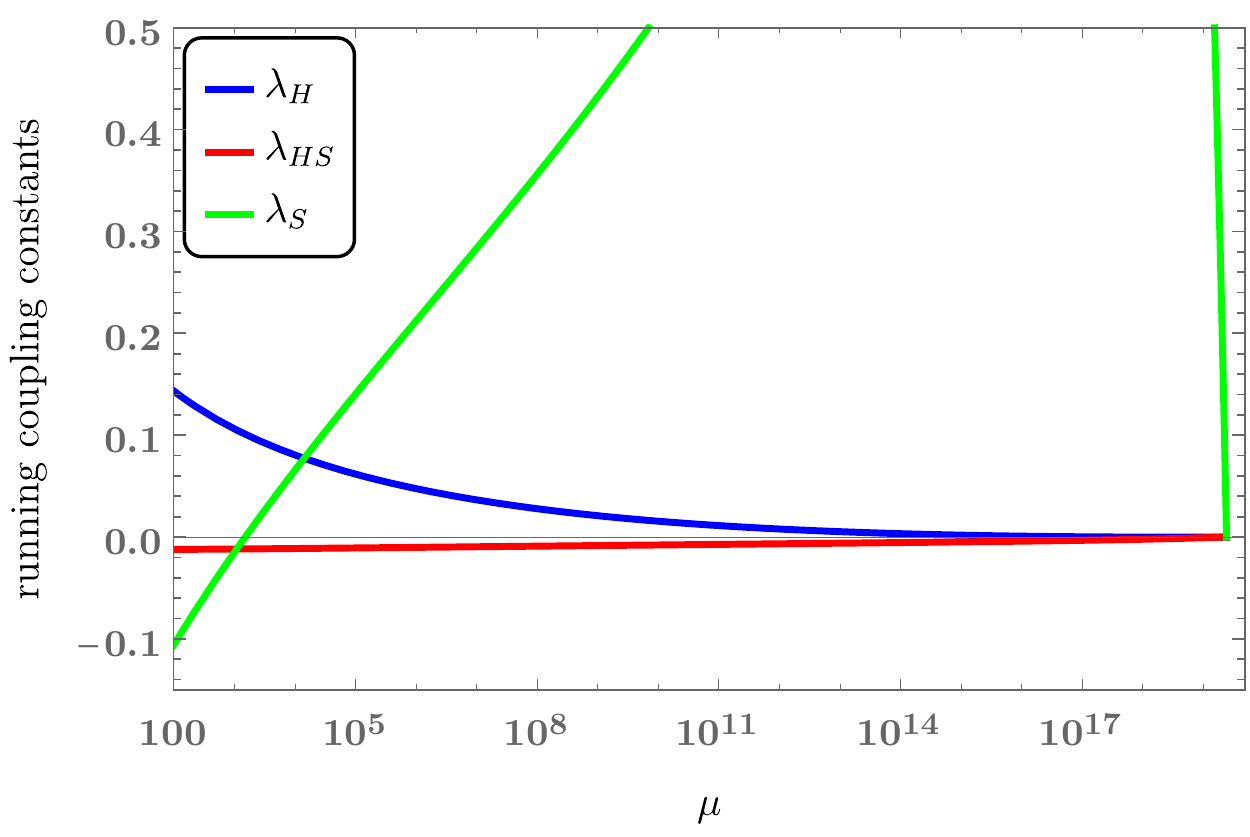}
\caption{
RG flow of scalar coupling constants from the Planck scale to the electroweak scale with the boundary condition \eqref{boundary condition for scalar interactions} and the benchmark value \eqref{benchmark values of couplings}.
}
\label{fig: RG flows} 
\end{figure}

We solve the RG equations for the system with the boundary conditions \eqref{boundary condition for scalar interactions} and \eqref{boundary condition for scalar mass}.
In Fig.\,\ref{fig: RG flows}, we show an example of the RG flows for the quartic and the Higgs portal coupling constants, where the following initial benchmark value is used:
\al{
&y_L\fn{M_\text{pl}}=1.842\,,&
&y_R\fn{M_\text{pl}}=1.354\,,&\nn[1ex]
&g_X\fn{M_\text{pl}}=0.794\,,&
&g_\text{mix}\fn{M_\text{pl}}=0.134\,.&
\label{benchmark values of couplings}
}
The boundary condition for the gauge coupling constants for the the SM gauge fields is given in Appendix~\ref{BC for gauge couplings}.
The Yukawa coupling constants $y_{L,R}$, the U(1)$_X$ gauge coupling constant $g_X$, and the kinetic mixing effect $g_\text{mix}$ are approximately constant within the RG flow since their beta functions are proportional to themselves.
One can see that the quartic coupling of $S$ is generated as a positive value in high energy region and then turns to a negative values at a certain renormalization scale. 
Such a behavior implies an occurrence of radiative symmetry breaking.
The Higgs portal coupling constant is generated as a negative value.

In order to extract information about the vacuum $v_S$, we consider the effective potential for $S$.
To this end, we here parametrize the complex scalar field $S$ such that 
\al{
S=\frac{\phi+i\eta}{\sqrt{2}}\,,
}
and then $v_S=\sqrt{2}\langle S\rangle =\langle \phi\rangle$.
Since the Higgs portal coupling constant is much smaller than the Yukawa coupling constant and the U(1)$_X$ gauge coupling constant, it could be negligible for the analysis of the vacuum.
This treatment allows us to consider the effective potential only for $\phi$.
Thus, the improved effective potential~\cite{Bando:1992np} for $\phi$ is given by
\al{
V_\text{eff}\fn{\phi}=\frac{\lambda_S\fn{t}}{4}G^4\fn{t}\phi^4\,,
}
where $\lambda_S$ is the running coupling constant obeying its RG equation, the dimensionless RG scale is parametrized as $t=\ln\fn{\phi/M}$ with a renormalization scale $M$.
Here, we choose $M=v_S$ for which $t=0$ corresponds to $\phi=v_S$.
The effect of the field renormalization is
\al{
G\fn{t}=\exp\left[ -\int^t_0 \df t'\,\gamma_S\fn{t'} \right]\,,
\label{field renormalization factor}
}
with the anomalous dimension of $S$,
\al{
\gamma_S\fn{t}=-\frac{\df \ln G}{\df t}=\frac{1}{32\pi^2}\left[ y_R^2+y_L^2 -24 g_X^2 \right]\,.
}

The vacuum $v_S$ is obtained from the stationary condition,
\al{
\phi\frac{\df V_\text{eff}}{\df \phi}\bigg|_{\phi=v_S}=\frac{\df V_\text{eff}}{\df t}\bigg|_{t=0}=0\,,
}
which gives a condition among coupling constants:
\al{
\bigg[4(1-\gamma_S)\lambda_S +\beta_{\lambda_S}\bigg]_{t=0}=0\,,
\label{condition for vS}
}
where 
\al{
\beta_{\lambda_{S}}=\frac{\df \lambda_S}{\df t}\,.
}
One can read the U(1)$_X$ breaking scale $v_S$ as a scale at which the condition \eqref{condition for vS} is satisfied.

After the U(1)$_X$ symmetry breaking, the singlet-scalar obtains a finite positive mass squared,
\al{
M_S^2\fn{v_S}=\frac{\df^2 V_\text{eff}}{\df \phi^2}\bigg|_{\phi=v_S}\simeq -4 \lambda_{S}\fn{v_S}v_S^2\,,
\label{mass for S}
}
where we neglect the running effect of the field renormalization \eqref{field renormalization factor} in the second equality.  
Furthermore, the effective quartic coupling constant is obtained by
\al{
\lambda_{\text{eff}\,S^4}=\frac{\df^4 V_\text{eff}}{\df \phi^4}\bigg|_{\phi=v_S}\simeq -\frac{22}{3}\lambda_S\fn{v_S}\,,
}
for which the singlet-scalar mass squared \eqref{mass for S} is
\al{
M_S^2\fn{v_S}=\frac{6}{11}\lambda_{\text{eff}\,S^4} v_S^2\,.
}
Note that the effective cubic coupling constant is given by
\al{
\lambda_{\text{eff},\,S^3}=\frac{\df^3 V_\text{eff}}{\df \phi^3}\bigg|_{\phi=v_S}\simeq -\frac{40}{3}\lambda_S\fn{v_S}v_S\,.
\label{Eq: cubic coupling}
}

A negative Higgs mass parameter are generated through a negative Higgs portal coupling such that at the renormalization scale $\mu=v_S$,
\al{
m_H^2\fn{v_S}=\frac{1}{2}\lambda_{HS}\fn{v_S}v_S^2\,.
\label{generated Higgs mass squared}
}
This mass parameter evolves until the electroweak scale $v_H$ by following its RG equation:
\al{
\mu\frac{\df m_H^2}{\df \mu}=\gamma_{m_H^2}\,,
}
where the anomalous dimension for the Higgs mass parameter $\gamma_{m_H^2}$ is given in Eq.\,\eqref{anomalous dimension for scalar masses}.

Let us turn to the Higgs sector with the generated Higgs mass parameter \eqref{generated Higgs mass squared}.
For a negligibly small Higgs portal coupling constant, the effective potential for the Higgs field is given by
\al{
V_\text{eff}\fn{h}=\frac{1}{2}m_H^2\fn{v_H}h^2+\frac{1}{4}\lambda_H\fn{v_H}h^4\,,
}
The electroweak vacuum is defined as the stationary point for the effective potential, $\df V_\text{eff}/\df h|_{h=v_H}=0$, for which one infers
\al{
v_H=\sqrt{-\frac{m_H^2\fn{v_H}}{\lambda_H\fn{v_H}}}\,.
}
At this vacuum, the Higgs boson mass is given by
\al{
M_H^2\fn{v_H}=\frac{\df^2V_\text{eff} }{\df h^2}\bigg|_{h=v_H}=2\lambda_H\fn{v_H}v_H^2+\Delta M_H^2\,,
}
where the last term on the right-hand side denotes the Higgs self-energy correction whose approximate form at the one-loop level is given by~\cite{Degrassi:2012ry,Kubo:2015joa}
\al{
\Delta M_H^2\simeq 16 C_{0} v_{H}^{2}\,,
}
with
\al{
C_{0} \simeq \frac{1}{64 \pi^{2} v_{h}^{4}}\left(3 M_{W}^{4}+(3 / 2) M_{Z}^{4}+(3 / 4) M_{h}^{4}-6 M_{t}^{4}\right)\,.
}
This correction can be obtained by using the one-loop effective potential given in Appendix~\ref{App: One-loop effective potential} and would be about 10\% within the physical Higgs mass.

So far, we have neglected the Higgs portal coupling.
This may be a good approximation for obtaining the physical mass spectra as long as a Higgs portal coupling is small.
Nevertheless, the mixing effect between the Higgs field and the singlet-scalar field plays a crucial role for the dark matter annihilation via these scalar fields.
The quadratic terms for the $(h,\phi)$-basis in the effective potential is diagonalized such that
\al{
\mathL_2&=
-\pmat{h & \phi}\pmat{
M_H^2 & M_{HS}^2 \\[1ex]
M_{HS}^2 & M_S^2
}\pmat{
h\\
\phi
}\nn[2ex]
&=-\pmat{ h' & \phi'}\pmat{
M_{h'}^2 & 0\\[1ex]
0 & M_{\phi'}^2
}\pmat{
h'\\
\phi'
}\,,
}
where we define $M_{HS}^2=\lambda_{HS} v_Hv_S$.
The mass eigenstates are given by 
\al{
\pmat{
h'\\
\phi'
}=
\pmat{
\cos \theta & -\sin \theta\\
\sin\theta & \cos\theta
}
\pmat{
h\\
\phi
}\,,\label{Eq: mixing matrix}
}
with the mass eigenvalues,
\al{
M_{h',\phi'}^2&= \frac{M_H^2 + M_S^2\pm \sqrt{(M_H^2 - M_S^2)^2+4M_{HS}^4}}{2}\nn
&=\frac{M_H^2 + M_S^2\pm (M_H^2 - M_S^2)\sqrt{1+\tan^22\theta}}{2}\,,
\label{eigen values of masses}
}
and the mixing angle,
\al{
\tan 2\theta=-\frac{2M_{HS}^2}{M_H^2 - M_S^2}\,.
\label{mixing angle}
}
For $M_H^2>M_S^2$ ($M_H^2<M_S^2$), the mixing angle is positive (negative).
The mixing angle between the Higgs and a new scalar boson is constraint such that $|\sin \theta| <0.3$~\cite{Martin-Lozano:2015dja,Falkowski:2015iwa}.
The Higgs mass in Eq.\,\eqref{eigen values of masses} has to satisfy the observed mass \eqref{observed masses and vacuum}, i.e. $M_{h'}=M_H^\text{obs}$.

In the mass eigenstates $(h',\phi')$, their propagators take forms
\al{
&\Delta_{h'h'}\fn{p^2}= \frac{1}{p^2-M_{h'}^2+i\Gamma_{h'} M_{h'}}\,,\nn[1ex]
&\Delta_{\phi'\phi'}\fn{p^2}= \frac{1}{p^2-M_{\phi'}^2+i\Gamma_{\phi'} M_{\phi'}}\,,
}
where $\Gamma_H$ and $\Gamma_S$ are decay widths for the Higgs boson and the scalar field $S$.
Using the mixing matrix \eqref{Eq: mixing matrix}, one obtains the propagators in the flavor basis $(h,\phi)$ so that
\al{
&\pmat{
\Delta_{HH} & \Delta_{HS}\\[1ex]
\Delta_{SH} & \Delta_{SS}
}=\nn[1ex]
&
\scriptsize{
\pmat{
\cos^2 \theta \, \Delta_{h'h'}+\sin^2 \theta \, \Delta_{\phi'\phi'} & \cos\theta \sin\theta (\Delta_{\phi'\phi}-\Delta_{h'h'} ) \\[2ex]
\cos\theta \sin\theta (\Delta_{\phi'\phi'}-\Delta_{h'h'}) & \cos^2 \theta \, \Delta_{\phi'\phi'}-\sin^2 \theta \, \Delta_{h'h'}
}
}\,.
\label{Eq: mixed propagators of scalar fields}
}

For the benchmark value of the coupling constants \eqref{benchmark values of couplings}, we obtain the expectation value of $S$,
\al{
v_S=1756.2~\text{[GeV]}\,,
}
for which we observe
\al{
&M_L=1114.3~\text{[GeV]}\,,
\quad
M_R=1042.5~\text{[GeV]}\,,\nn[2ex]
&M_X=1380.8~\text{[GeV]}\,,
\quad
M_{\phi'}=227.9~\text{[GeV]}\,.
}

\subsection{Decay of new particles}
\label{subsec: Decay of scalar fields}
We note decay processes of the Higgs field and the singlet-scalar field.
Due to the mixing between the Higgs field and the singlet-scalar field, the partial decay width of the singlet-scalar and the Higgs fields into the SM particles are given by
\al{
&\Gamma_{\phi'}\fn{\phi'\to \text{SMs}}=\sin^2\theta \times \Gamma\fn{h_\text{SM}\to \text{SMs}}|_{\small M_{h'}\to M_{\phi'}}\,,\\[2ex]
&\Gamma_{h'}\fn{h'\to \text{SMs}}=\cos^2\theta \times \Gamma\fn{h_\text{SM}\to \text{SMs}}\,,
}
respectively, where the right-hand side is the partial decay width of the Higgs evaluated in the SM.
For a small Higgs portal coupling constant, there is no significant deviation of $\Gamma_H$ from the SM case and the partial decay width of the singlet-scalar field is small.
Since our model predicts $v_S>v_H$, the Majorana fermions are heavier than the Higgs boson, and then the Higgs field does not decay into them.
On the other hand, if $2M_{\phi'}<M_{h'}$ the decay channel $h'\to \phi'\phi'$ opens:
\al{
\label{eq: decay width of Higgs}
\Gamma_{h'}\fn{h'\to \phi'\phi'}=\frac{(\lambda_{HS}v_H)^2}{8\pi M_h^2}\sqrt{\frac{M_{h'}^2}{4}-M_{\phi'}^2}\,.
}

The extra U(1)$_X$ boson decays into SM particles via the kinetic mixing effect with $g_\text{mix} \gsim 10^{-9}$.
This bound is adequately small for the extra U(1)$_X$ boson to decay in the current our system.
Thus, the Majorana fermions annihilate into $X_\mu$ if $M_{R,L}>M_X$.

\subsection{Allowed parameter space}
We scan the parameter space where the phenomenological constrains \eqref{observed masses and vacuum} and \eqref{observed relic abundance} are satisfied.
We here summarize constraints for free parameters in our model.
As can be seen in Eq.\,\eqref{observed masses and vacuum}, the top-quark mass has a large uncertainty. We perform the parameter search with a fixed value $M_t\simeq 160.4$\,GeV for which the Higgs mass could be generated so as to $M_{h'}\simeq 125$\,GeV within the SM. Although the extension of the SM could give a small deviation of the Higgs mass, it is still consistent within the uncertainty of the top-quark mass.
From the asymptotic safety condition above the Planck scale, the gauge coupling constant should satisfy the bound $g_X\fn{M_\text{pl}}\lsim 1.09$ and $g_\text{mix}\fn{M_\text{pl}}\lsim 0.68$.
To realize the scalegenesis in the flatland scenario, the ratio $r=y_{L,R}/g_X$ should be in the range $1.04\lsim r\fn{v_S} \lsim 1.32$ which gives a constraint for the Yukawa coupling constants by combing with the bound for the gauge coupling constant $g_X$.

\begin{figure}
\includegraphics[width=8.5cm]{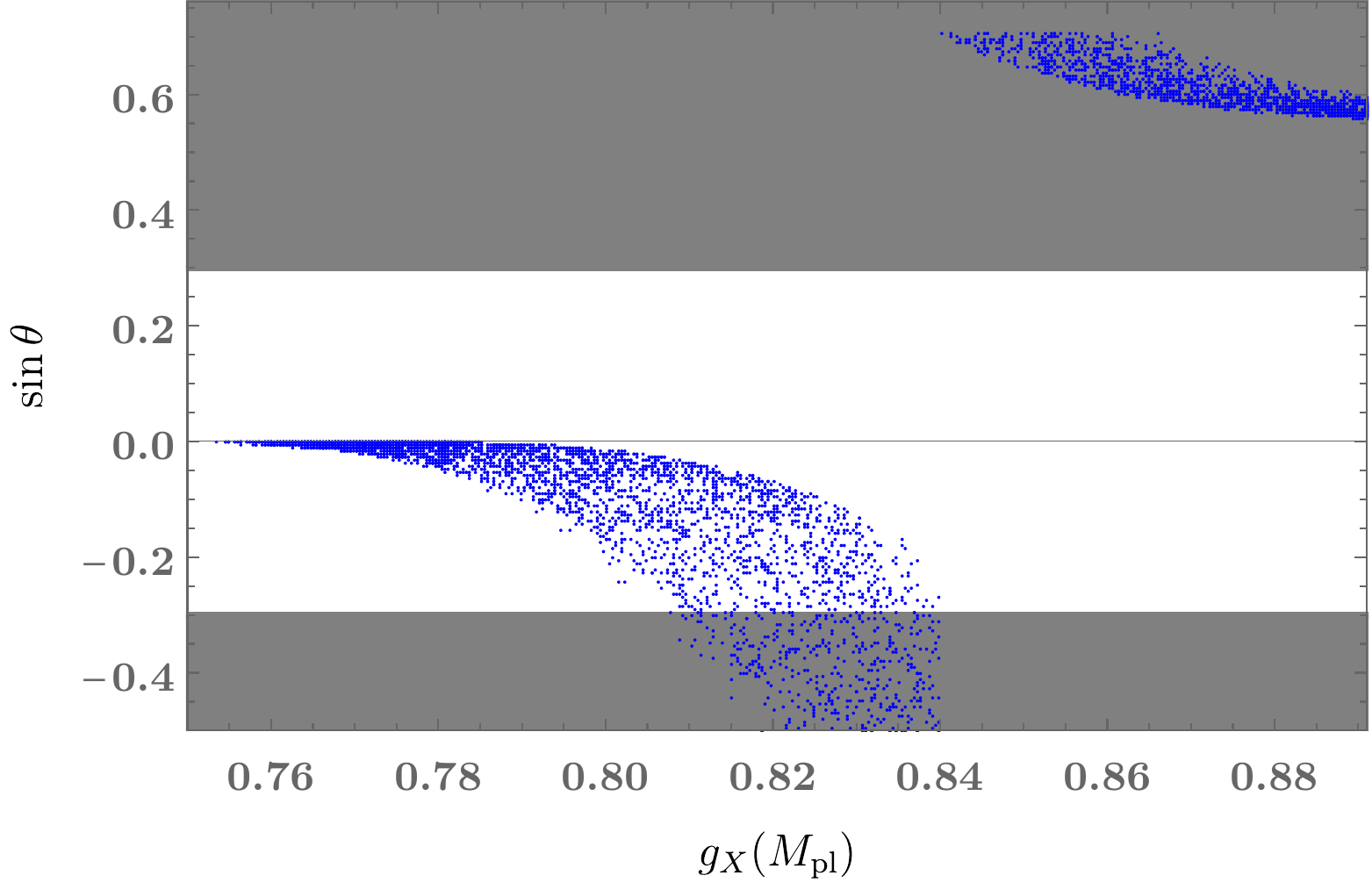}
\caption{
Region plot for the mixing angle, $\sin\theta$, as a function of $g_X\fn{M_\text{pl}}$.
The grey shadow region ($|\sin\theta|>0.3$) is excluded by the collider experiments~\cite{Falkowski:2015iwa}.
}
\label{plot: mixing angle as a function of gX} 
\end{figure}
We first show the mixing angle ($\sin\theta$) as a function of $g_X(M_\text{pl})$.
Fig.\,\ref{plot: mixing angle as a function of gX} represents the region for $\sin\theta$ where the constraints above are satisfied. 
The grey shadow region $|\sin\theta|<0.3$ is excluded by collider experiments~\cite{Falkowski:2015iwa}.
One can see from Fig.\,\ref{plot: mixing angle as a function of gX} that for $g_X(M_\text{pl})\gsim 0.84$ the value of $\sin\theta$ is positive, i.e., the singlet-scalar boson is lighter than the Higgs boson, and this region is already excluded.
Therefore, hereafter we restrict the value of the U(1)$_X$ gauge coupling to $g_X(M_\text{pl})\lsim 0.84$.

\begin{figure*}
\includegraphics[width=8.5cm]{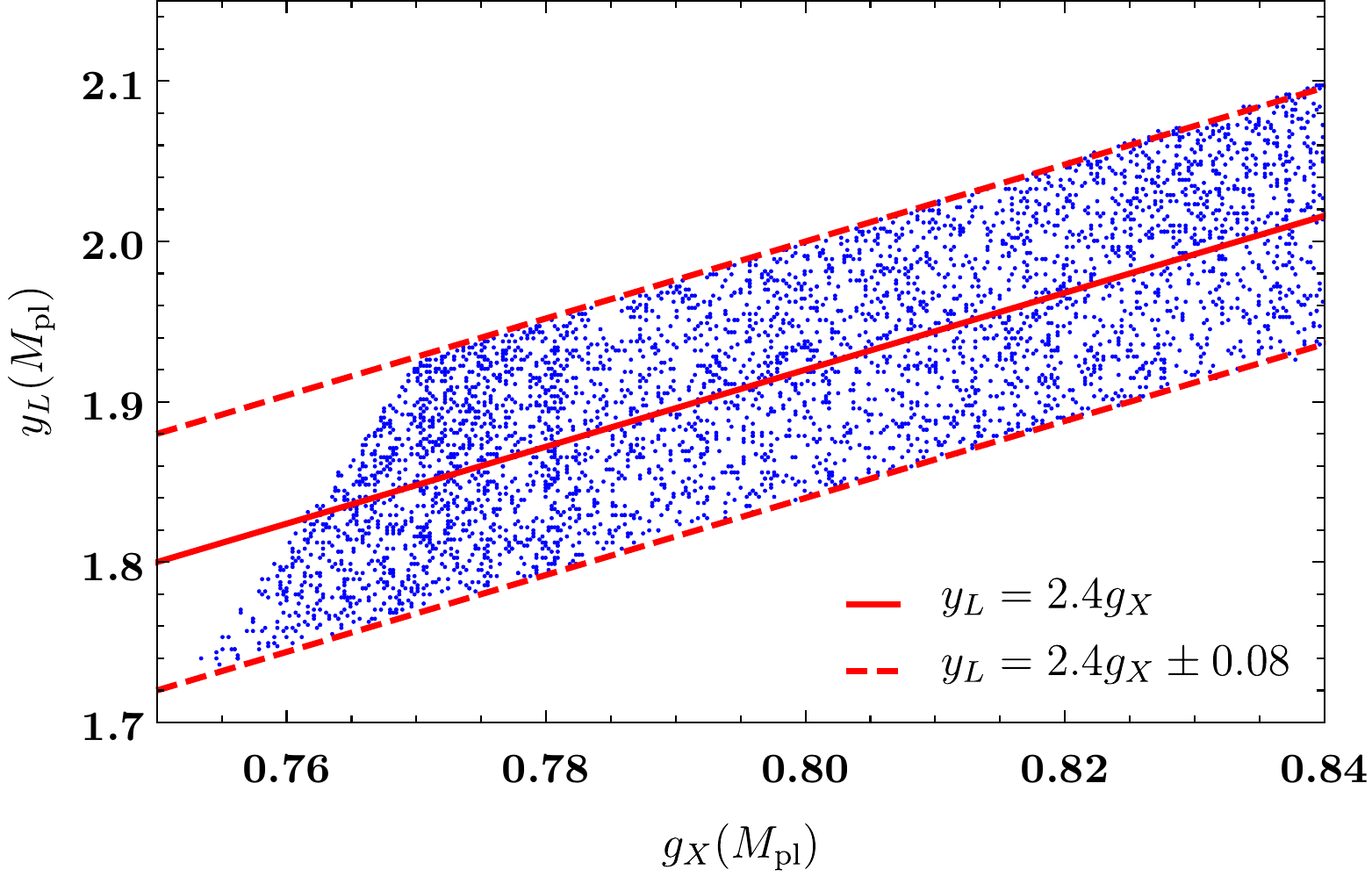}
\includegraphics[width=8.5cm]{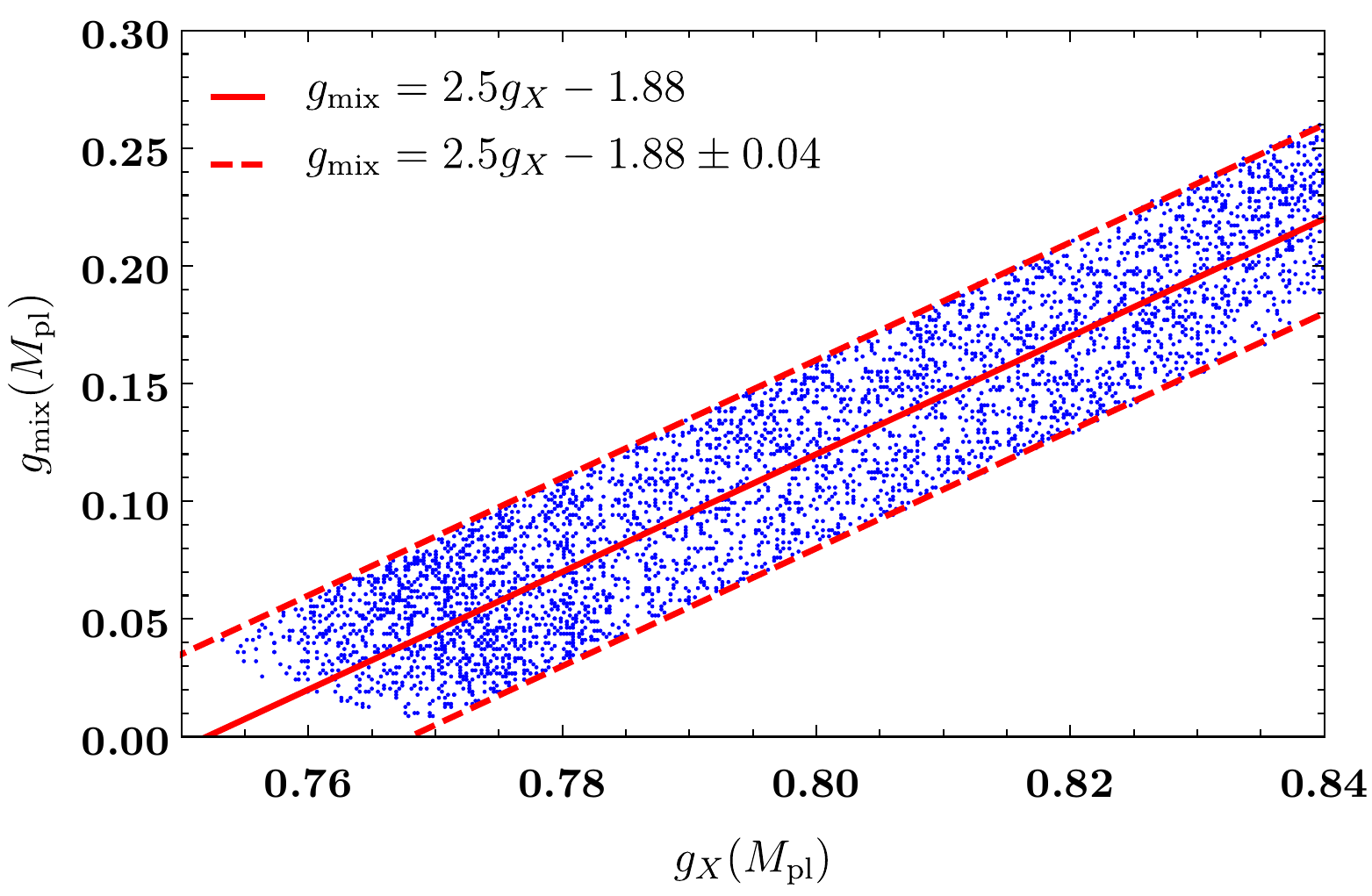}
\caption{
Allowed region plots for coupling constants, $y_{L}(M_\text{pl})$ (left) and $g_\text{mix}(M_\text{pl})$ (right), as functions of $g_X\fn{M_\text{pl}}$.
}
\label{plot: physical coupling quantities as a function of gX} 
\end{figure*}

In Fig.\,\ref{plot: physical coupling quantities as a function of gX} we show $y_L(M_\text{pL})$ and $g_\text{mix}(M_\text{pL})$ as functions of $g_X(M_\text{pl})$.
These couplings have a linear dependence on the U(1)$_X$ gauge coupling.
In Fig.\,\ref{plot: physical coupling quantities as a function of gX} we plot linear red lines which are given by, at the Planck scale,
\al{
&y_L\simeq 2.4g_X\,,&
&g_\text{mix}\simeq 2.5g_X-1.8\,.&
\label{Eq: relation between yL, gmix and gX}
}
We parametrize the the right-handed Yukawa coupling as
\al{
y_R(M_\text{pl})=\xi y_L(M_\text{pl})\,,
\label{Eq: relation between yL and yR}
}
where $\xi$ is a constant less than 1.
With Eqs.\,\eqref{Eq: relation between yL, gmix and gX} and \eqref{Eq: relation between yL and yR}, $g_X(M_\text{pl})$ and $\xi$ can be cast as free parameters in this system.
One of them will be constrained such that the relic abundance of the Majorana fermions satisfy the dark matter relic abundance.

\begin{figure*}
\includegraphics[width=8.5cm]{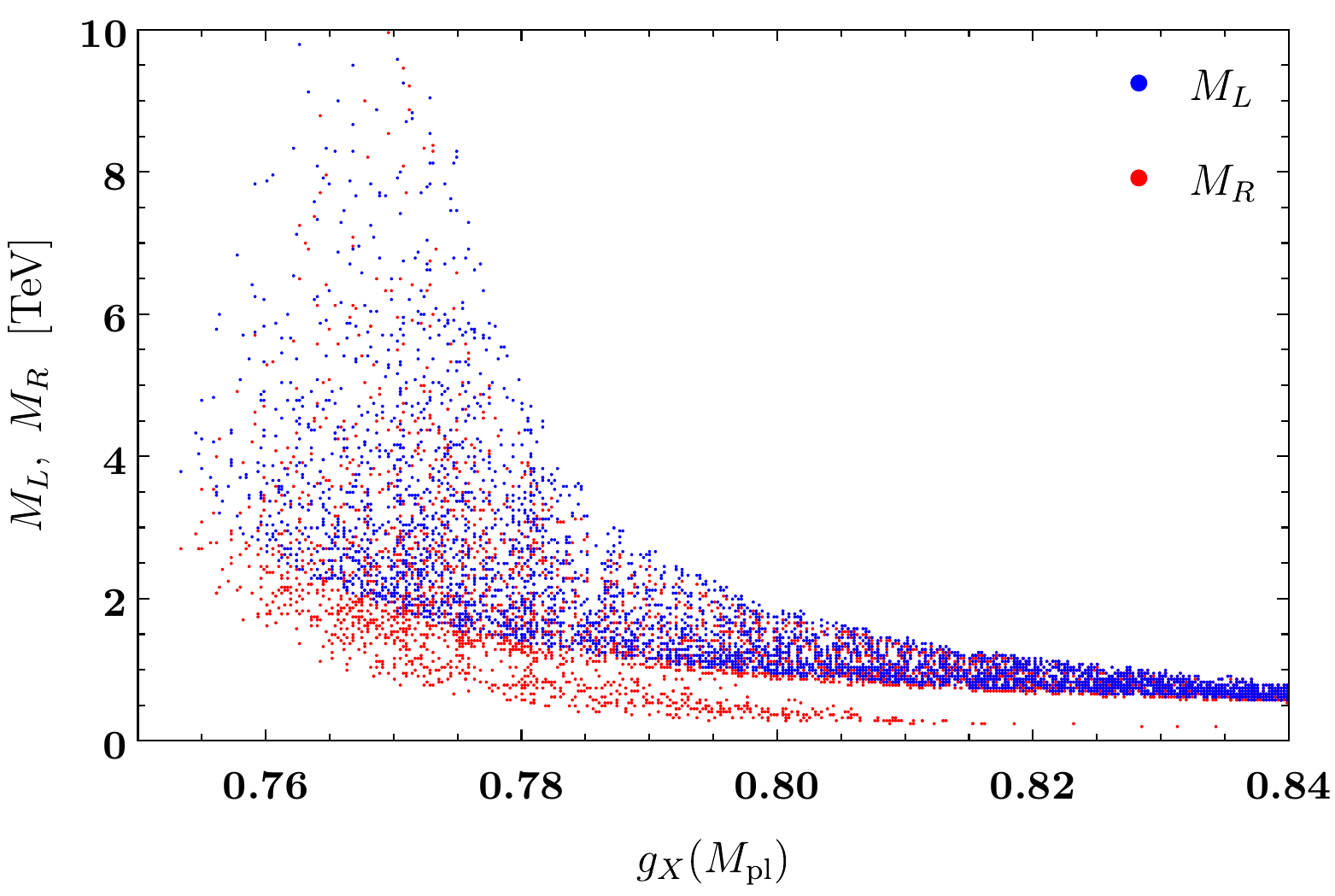}
\includegraphics[width=8.5cm]{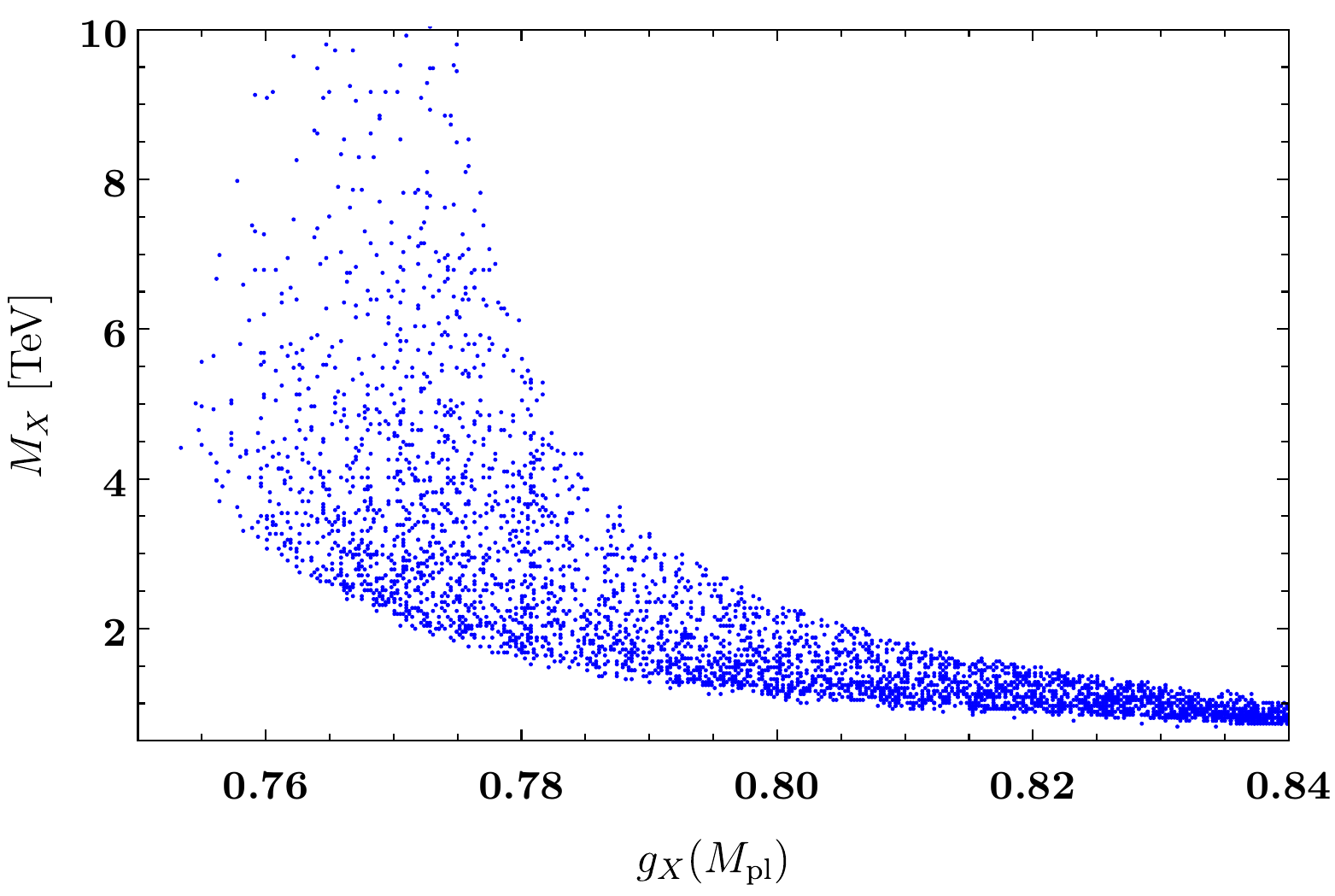}
\includegraphics[width=8.5cm]{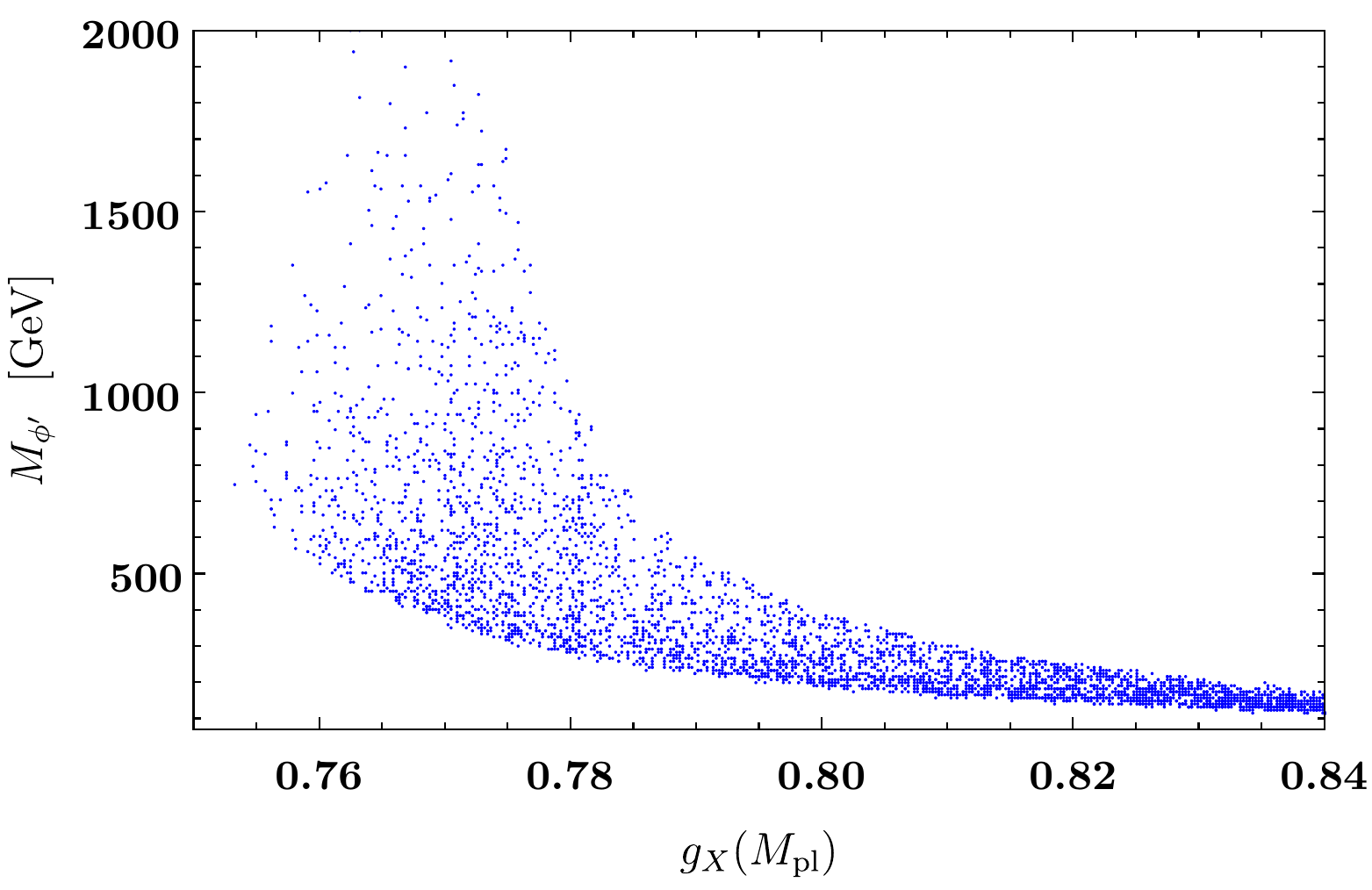}
\includegraphics[width=8.5cm]{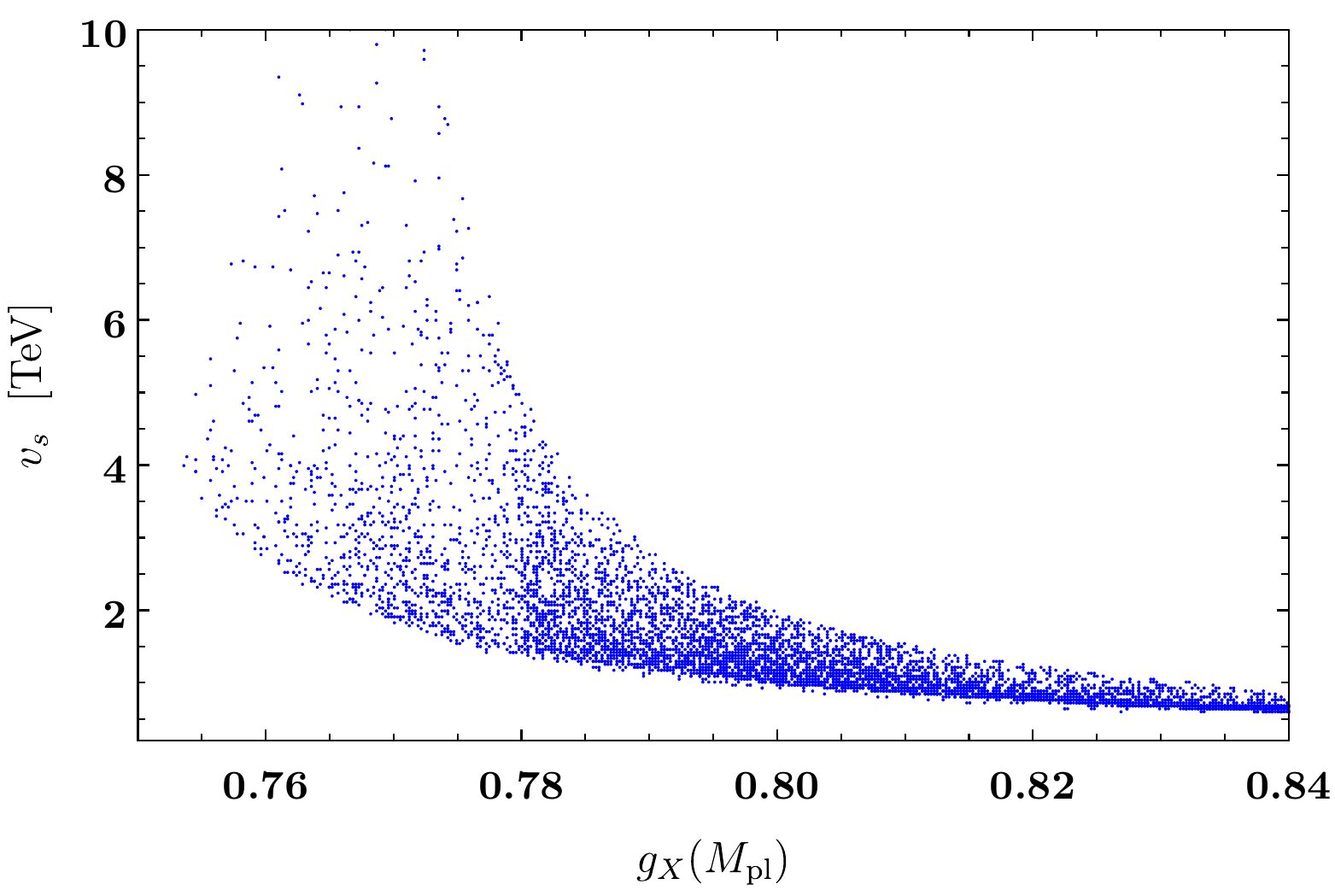}
\caption{
Allowed region plots for physical quantities, i.e. $M_{L,R}$, $M_X$, $M_{\phi'}$ and $v_S$ as functions of $g_X\fn{M_\text{pl}}$.
}
\label{plot: physical dimensionful quantities as a function of gX} 
\end{figure*}

Fig.\,\ref{plot: physical dimensionful quantities as a function of gX} exhibits allowed region for dimensionful physical quantities ($M_L$, $M_R$, $M_X$, $M_{\phi'}$ and $v_S$).
One can see from Fig.\,\ref{plot: physical dimensionful quantities as a function of gX} that these quantities tend to be in inverse proportion to $g_X(M_\text{pl})$.

\section{Majorana fermions as dark matter candidates}
\label{Dark matter analysis}
In this section we investigate the properties of the Majorana fermions as dark matters.
We start by setting up the Boltzmann equation to evaluate the relic density of the Majorana fermions within the cosmological evolution, and then the allowed parameter region, where the the observed relic density is satisfied, is searched.

\subsection{Boltzmann equation and dark matter relic density}
There are two dark matter candidates, namely $\chi_R$ and $\chi_L$.
In order to follow evolutions of their number densities $n_{R,L}$ as functions of temperature, we here introduce the Boltzmann equations.
Since the structure of the Boltzmann equations is symmetric under the exchange $L\leftrightarrow R$, we here show the case only for the left-handed side.
Instead of the number densities, it is useful to introduce the quantities $Y_{R,L}=n_{R,L}/s$, where $s$ is the entropy density.
The Boltzmann equation for $Y_L$ is given by~\cite{DEramo:2010keq,Belanger:2011ww,Belanger:2012vp,Aoki:2012ub,Kubo:2017wbv}
\al{
\frac{\df Y_L}{\df x}&=-0.264g_*^{1/2}\left[ \frac{\mu_{RL} M_\text{pl}}{x^2}\right]\nn
&\quad\times\Bigg[\langle \sigma\fn{\chi_L\chi_L;\text{SMs, $\phi$, $X_\mu$}} v\rangle \left(Y_L^2-\overline{Y}_L^2\right) \nn
&\quad
+\langle \sigma\fn{\chi_L\chi_L;\chi_R\chi_R} v\rangle\left(Y_L^2-\frac{Y_R^2}{\overline{Y}_R^2}\overline{Y}_L^2  \right)
\Bigg]\,,
}
where $M_\text{pl}=1.22\times 10^{19}$\,GeV is the Planck mass; $g_*=106.75$ is the total number of effective degrees of freedom in the SM; $1/\mu_{RL}=1/M_R+1/M_L$ is the reduced mass; $x=\mu_{RL}/T$ is the dimensionless inverse temperature; and $\overline{Y}_L$ is $Y_L$ in the thermal equilibrium, 
\al{
\overline{Y}_L\fn{x}=\frac{45x^2}{4\pi^4g_{*}}\frac{M_L^2}{\mu_{RL}^2}K_2\fn{(M_L/\mu_{RL})x}\,,
}
with $K_2\fn{x}$ the modified Bessel function of the second kind.
Here, $\langle \sigma\fn{\chi_L\chi_L;\text{SMs, $\phi$, $X_\mu$}} v\rangle$ is the thermal averaged cross section for the dark matter annihilation processes.
Such annihilations take place with mediators, the singlet-scalar field $\phi$~\cite{Kainulainen:2015sva,Krnjaic:2015mbs,Matsumoto:2018acr}, the Higgs field $h$, the U(1)$_X$ gauge field $X_\mu$~\cite{Brahmachari:2014aya} and the Majorana fermion as exhibited in Fig.\,\ref{dark matter annihilation processes} in Appendix~\ref{App: Cross sections for dark matter annihilation}.
As discussed in subsection~\ref{subsec: Decay of scalar fields}, the singlet-scalar fields in the final state decay into lighter SM particles through the interaction with the Higgs field.
These mediators cause also the $\chi_L\chi_L\to \chi_R\chi_R$ scattering of the Majorana fermions, whose thermal averaged cross section is denoted by $\langle \sigma\fn{\chi_L\chi_L;\chi_R\chi_R} v\rangle$.
These processes are show in Fig.\,\ref{dark matter annihilation processes LL->RR} in Appendix~\ref{App: Cross sections for dark matter annihilation} where we show their explicit forms.

Solving the coupled Boltzmann equation for $Y_R$ and $Y_L$, one can evaluate the relic density of the dark matter,
\al{ 
\Omega_\text{DM}\hat{h}^2
&=\Omega_R\hat{h}^2+\Omega_L\hat{h}^2\nn
&=\frac{s_0}{\rho_c/\hat{h}^2}\left( Y_{R,\infty} M_R +Y_{L,\infty} M_L\right),
\label{Eq: relic abundance of Majorana fermions}
}
where we have $s_0=2890\,\text{cm}^{-3}$ and $\rho_c/\hat{h}^2=1.05\times 10^{-5}\,\text{GeV}\,\text{cm}^{-3}$~\cite{Tanabashi:2018oca,Aghanim:2018eyx}, and $Y_{R,L,\infty}$ are the values of $Y_{R,L}$ at $x=\infty$ corresponding to the zero temperature.
In our working assumption $M_R<M_L$ (or equivalently $y_R<y_L$), the annihilation of $\chi_R$s to $\chi_L$s does not take place, namely $\langle \sigma\fn{\chi_R\chi_R;\chi_L\chi_L} v\rangle=0$.
The left-handed Majorana fermions annihilate to the right-handed ones in addition to the SM particles and the singlet-scalar bosons within the temperature evolution so that the main ingredient of the dark matter relic density is the right-handed Majorana fermions.

\begin{figure*}
\includegraphics[width=8.5cm]{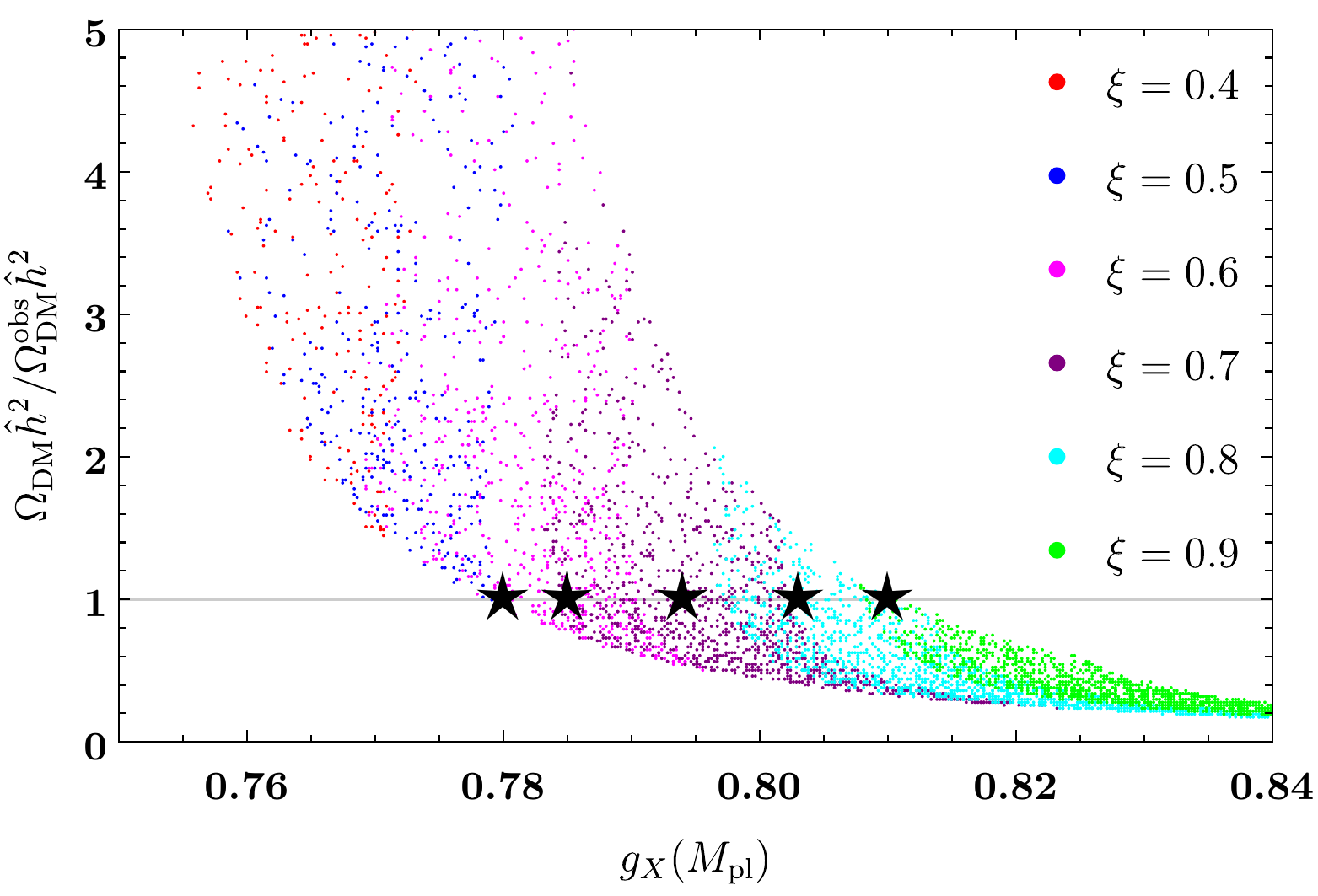}
\includegraphics[width=8.5cm]{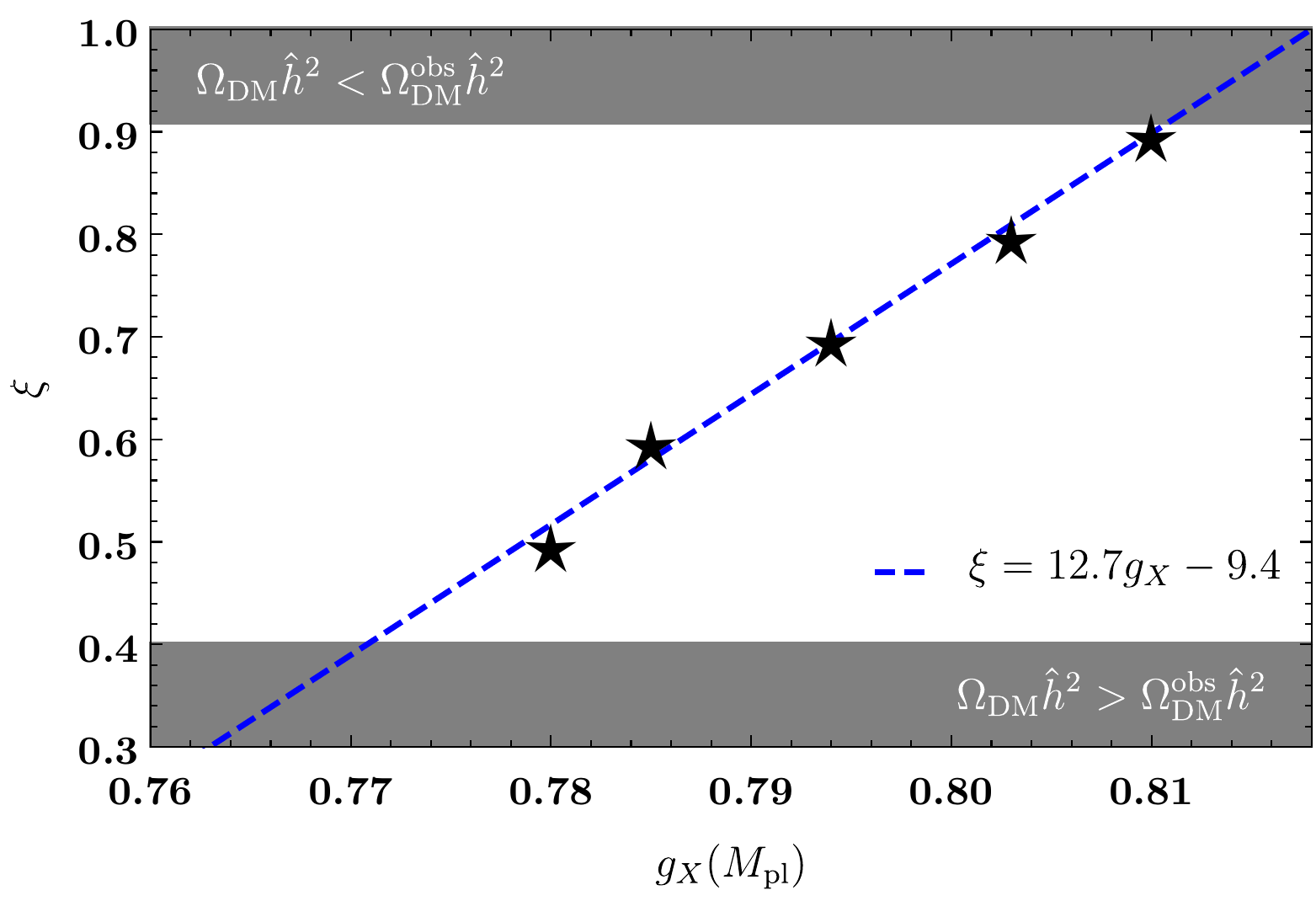}
\caption{
Left: The relic abundances of the Majorana fermions \eqref{Eq: relic abundance of Majorana fermions} normalized by the observed value \eqref{observed relic abundance} as a function of $g_X(M_\text{pl})$ for a fixed value of $\xi=y_R(M_\text{pl})/y_L(M_\text{pl})$. The star points are located at the typical values of $g_X(M_\text{pl})$ satisfying $\Omega_\text{DM}\hat{h}^2=\Omega_\text{DM}^\text{obs}\hat{h}^2$ for each value of $\xi$.
Right: $\xi$ and $g_X(M_\text{pl})$ at the star points. The blue dashed linear line is obtained by fitting to these points.
In the gray region, $\Omega_\text{DM}\hat{h}^2/\Omega_\text{DM}^\text{obs}\hat{h}^2=1$ is not satisfied. 
}
\label{plot: relic abundance and gX} 
\end{figure*}

The left-hand side panel of Fig.\,\ref{plot: relic abundance and gX} exhibits the relic abundance of the Majorana fermions \eqref{Eq: relic abundance of Majorana fermions} normalized by the observed one \eqref{observed relic abundance} as a function of $g_X(M_\text{pl})$.
There exists a region satisfying $\Omega_\text{DM}\hat{h}^2=\Omega_\text{DM}^\text{obs}\hat{h}^2$ in between $0.78\lsim g_X(M_\text{pl})\lsim 0.81$.
The star points denote typical points for each value of $\xi=y_R(M_\text{pl})/y_L(M_\text{pl})$.
We show the values (star points) of $\xi$ as a function of $g_X(M_\text{pl})$ in the right-hand side panel of Fig.\,\ref{plot: relic abundance and gX}.
One can see the linear dependence of $\xi$ on $g_X(M_\text{pl})$.
The linear fitting yields the relation for $0.78\lsim g_X(M_\text{pl})\lsim 0.81$,
\al{
\xi\simeq  	12.7g_X(M_\text{pl})-9.4\,.
\label{Eq: xi and gX relation}
}
Then, all parameters excepts for $g_X(M_\text{pl})$ are fixed by the observed data.

\subsection{Prediction on Spin-independent elastic cross section}
\begin{figure}
\includegraphics[width=6cm]{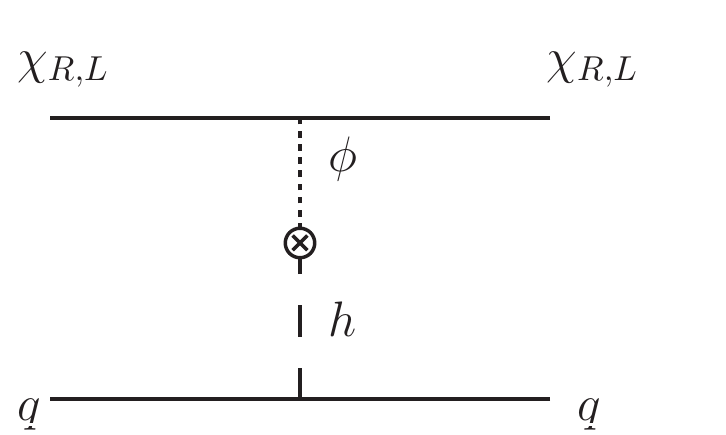}
\caption{
The $\chi_{L,R}$-quark scattering process yielding the spin-independent elastic cross section of a dark matter.
The kinetic mixing is denoted by the cross.
The cross in a circle stands for the mixing between $\phi$ and $h$.
}
\label{dark matter quark scattering} 
\end{figure}

In the WIMP dark matter search, interactions between a nucleon and a dark matter play a crucial role for the detection of a dark matter signal.
They could be observed as the spin-independent (SI) elastic cross section of a dark matter and a nucleon~\cite{Barbieri:2006dq,Aoki:2012ub}.
In our model, the scattering of the Majorana fermions and quarks could take place as the $t$-channel diagram in the processes displayed in Fig.\,\ref{dark matter quark scattering} from which one can obtain the effective scalar-type four-Fermi interaction $\mathL_\text{eff}=G_q^{R}(\overline{\chi_{R}^c}\chi_{R})(\bar q q)+G_q^{L}(\overline{\chi_{L}^c}\chi_{L})(\bar q q)$.
More specifically, the coefficient of the four-Fermi interactions are calculated as
\al{
\label{Eq: effective four-fermi coupling}
G_q^{R,L}&= \frac{y_qy_{R,L}}{2}\cos\theta\sin\theta\left( \frac{1}{M_\phi^2}-\frac{1}{M_h^2}\right)\,,
}
with $y_q$ a quark Yukawa coupling constant.
Note that since the quark-DM interaction induced by the $X$ gauge boson exchange gives the spin-dependent cross section, we do not evaluate it in this work.
The four-Fermi interaction \eqref{Eq: effective four-fermi coupling} between a Majorana fermion and a quark is translated into the effective interactions between a Majorana fermion and a nucleon by
\al{
G_N^{R,L}=\sum_{q=\text{all quarks}} f_q^N G_q^{R,L}\frac{M_N}{M_q}\,,
}
where $M_q=y_q v_H/\sqrt{2}$ is a quark mass, $M_N\simeq 939$\,MeV is the nucleon mass, and $f_q^N=M_q\langle N| \bar q q|N\rangle/M_N$ is each quark matrix element of a nucleon.

For the left-handed Majorana fermion $\chi_L$, the SI elastic cross section of a dark matter and a nucleon is computed as
\al{
\sigma_{\text{SI}}^L
=\frac{4\mu_{NL}^2}{\pi}\Bigg[& \left(\frac{M_N  f_N y_L}{\sqrt{2}v_H} \right)\cos\theta \sin\theta \left( \frac{1}{M_\phi^2}-\frac{1}{M_h^2}\right)\Bigg]^2\,,
}
where $\mu_{NL}=M_NM_L/(M_N+ M_L)$ is the reduced mass for the $N$-$\chi_L$ system, and $f_N$ is calculated as
\al{
f_N&=\sum_{q=\text{all quarks}}f_q^N\simeq  0.305\,,
}
with $f_q^N$ evaluated in~\cite{Junnarkar:2013ac,Crivellin:2013ipa,Hoferichter:2015dsa}.
One can obtain the case for the left-handed Majorana fermion by exchanging $L\leftrightarrow R$.

\begin{figure*}
\includegraphics[width=9cm]{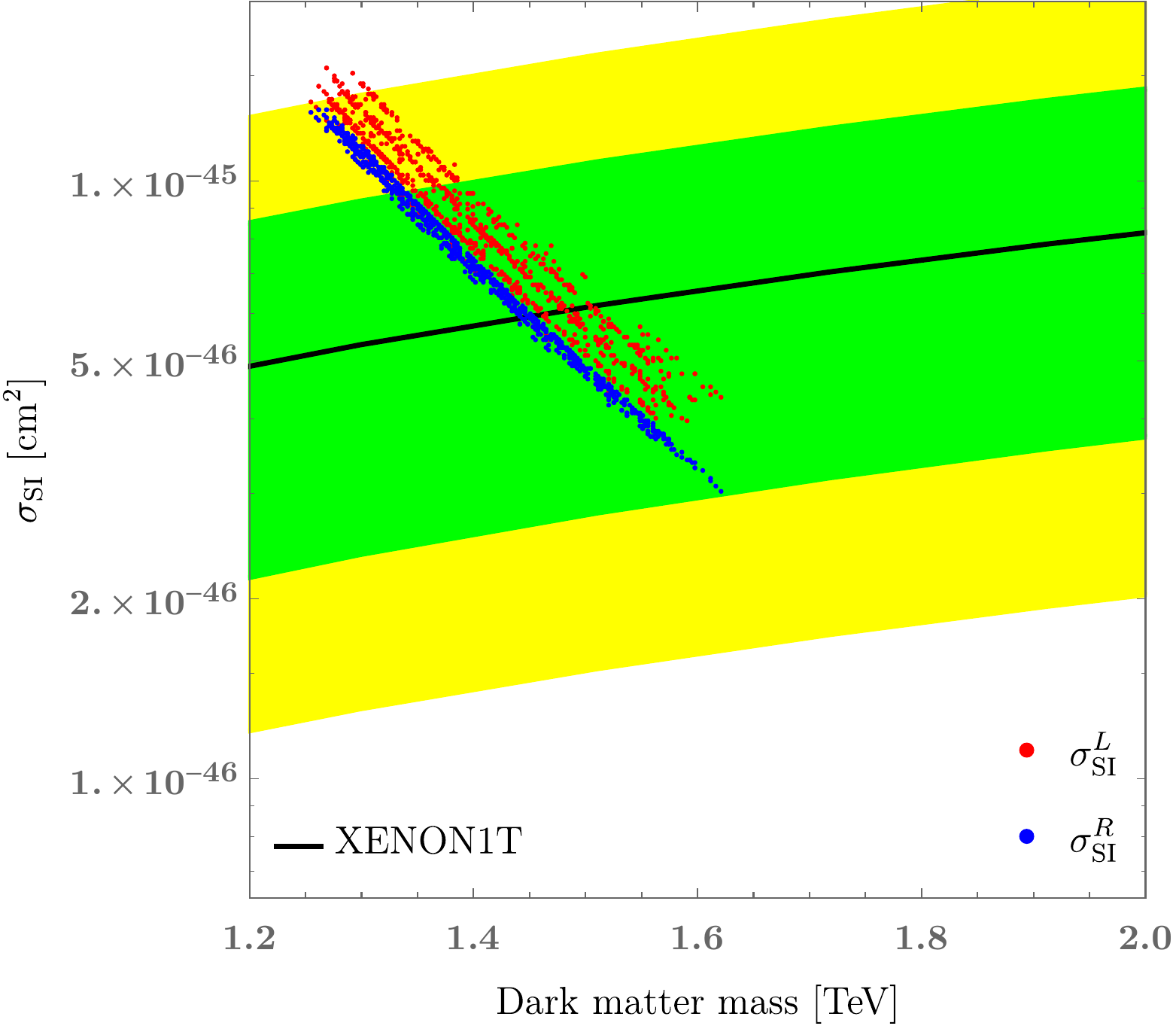}
\caption{
The SI elastic cross section of Majorana fermions as a function of their masses.
The black solid line is the current upper bound of XENON1T~\cite{Aprile:2018dbl}.
The green and yellow bands stand for the $1\sigma$ and $2\sigma$ bands, respectively.
}
\label{SI cross section} 
\end{figure*}

In Fig.\,\ref{SI cross section}, we plot the spin-independent elastic cross sections with the upper bound provided by XENON1T~\cite{Aprile:2018dbl}.
The constraints from LUX~\cite{Akerib:2016vxi} and PandaX-II~\cite{Cui:2017nnn} are somewhat milder than those of XENON1T; see~\cite{Aprile:2018dbl}.
One can see from Fig.\,\ref{SI cross section} that there is a small allowed region slightly below the upper bound (solid black line).

The model has the strong predictability thanks to the conditions from the asymptotically safe quantum gravity scenario.
If the spin-independent elastic cross section is observed, all parameters in the model are determined.
The Majorana fermions as dark matter candidates in the model could be tested in the near future.

\section{Summary}
\label{sect: summary}
We have proposed an extension of the SM based on the flatland scenario with dark matter candidates.
The flatland condition corresponds to the fact that all scalar interactions including mass terms vanish at the Planck scale, namely the scalar potential is flat above the Planck scale, especially the model is scale invariant. 
Such a condition could be compatible with the asymptotic safety program of quantum gravity.
We introduce Majorana fermions coupled to a U(1)$_X$ gauge field and a singlet-scalar field.
The U(1)$_X$ gauge field interacts with the U(1)$_Y$ gauge field in the SM even at the classical level via the kinetic mixing effect.
At this point, there are four free parameters ($g_X$, $g_\text{mix}$, $y_L$ and $y_R$).
This is a minimal setup of an extended model compatible with the flatland conditions which could be naturally concluded from the asymptotically safe quantum gravity scenario.
We have demonstrated that the minimal extension of the SM contains eventually only one free parameters and then has a strong predictability.

Let us here summarize the processes that the four parameters are fixed.
The condition for the electroweak scalegenesis due to the Coleman-Weinberg mechanism gives a constraint for the ratio between the Yukawa coupling and the  U$_X$(1) gauge coupling.
Hence, the electroweak scale $v_H=246$\,GeV fix one of the Yukawa couplings.
The kinetic mixing effect $g_\text{mix}$ generates a finite negative value of the Higgs portal coupling between the Higgs doublet-field in the SM and the singlet-scalar field, so that the observed Higgs mass $m_H=125$\,GeV determines the value of $g_\text{mix}$.
We found $y_L(M_\text{pl})$ and $g_\text{mix}$ as functions of $g_X$ by the numerical analysis, i.e. Eq.\,\eqref{Eq: relation between yL, gmix and gX}.
The relic abundance of the Majorana fermions has to be satisfied the current observed value \eqref{observed relic abundance}.
From this constraint, we determined the ratio between the Yukawa couplings, denoted by $\xi$, as given in Eq.\,\eqref{Eq: xi and gX relation}.
Then, there is only one free parameter, e.g. $g_X(M_\text{pl})$ in the model.

We have evaluated the SI elastic cross section of Majorana fermions and a nucleon.
The model predicts the SI elastic cross sections as functions of the Majorana masses around the current upper bound of XENON1T.
There is a small allowed region slightly below the upper bound.
Therefore, the Majorana fermions in the model as dark matter candidates could be tested by the direct detection experiments of the WIMP dark matter such as XENON, LUX and PandaX-II.
If the SI elastic cross section is observed, all parameters in the model are determined. 
Hence, it is important to investigate the possibilities of the observations of the other particles, i.e. the singlet-scalar boson mass and the U(1)$_X$ gauge boson mass. 
The future collider experiments such as the High-Luminosity Large Hadron Collider (HL-LHC)~\cite{Apollinari:2017cqg} and the International Linear Collider (ILC)~\cite{BrauJames:2007aa} could find these particles.

The investigation of stochastic gravitational waves produced by phase transitions at finite temperature may be one of other possible tests for the model.
It is actually reported in \cite{Jinno:2016knw} that a similar model (classically scale invariant $B-L$ model) can produce gravitational waves whose spectra could be tested by future interferometer experiments.
In such a case, one expects that a supercooling universe is realized. 
In particular, when the electroweak phase transition temperature is lowered until the QCD phase transitions, the thermal history of the universe could be changed drastically~\cite{Iso:2017uuu}.
It is interesting subject to investigate the nature of our model in a supercooling universe.

\subsection*{Acknowledgements}
We thank Manuel Reichert for valuable discussions.
Y.\,H. thanks the strongly correlated systems group of Institut f\"ur Theoretische Physik, Universit\"at Heidelberg for their kind hospitality.
The work of Y.\,H. is supported by a Grant-in-Aid for JSPS Fellows (No.\,JP18J22733).
The work of K.\,T. is supported by the MEXT Grant-in-Aid for Scientific Research on Innovation Areas (KAKENHI Grant Numbers No.\,JP18H05543).
The work of M.\,Y. is supported by the Alexander von Humboldt Foundation. 

\newpage
\onecolumngrid
\appendix
\section{Renormalization group equations}
\label{beta functions}
\subsection{Beta functions}
Here, we list the beta functions for coupling constants at the one-loop level in the extended model \eqref{starting action}.
The beta functions have similar structures to a $B-L$ extension of the SM~\cite{Basso:2010jm}.

For the gauge coupling constants in the SM sector, one has
\al{
\text{U(1)$_Y$}:&~~(4\pi)^2\beta_{g_Y}=\frac{41}{6}g_Y^3\,,\nn[1ex]
\text{SU(2)$_L$}:&~~(4\pi)^2\beta_{g_2}=-\frac{19}{6}g_2^3\,,\nn[2ex]
\text{SU(3)$_c$}:&~~(4\pi)^2\beta_{g_S}=-7g_S^3\,,
}
while the beta functions for the gauge coupling constant for the new gauge field $X_\mu$ and the kinetic mixing effect are given by
\al{
\label{App: beta functions for new gauge couplings}
&(4 \pi)^{2} \beta_{g_{X}}=\frac{8}{3} g_{X}^{3}+\frac{41}{6} g_{X} {g}_\text{mix}^{2}\,,\nn[2ex]
&(4 \pi)^{2} \beta_{{g}_\text{mix}}= \frac{41}{6} {g}_\text{mix}\left({g}_\text{mix}^{2}+2 g_{Y}^{2}\right)+\frac{8}{3} g_{X}^{2} {g}_\text{mix}\,.
}

The Yukawa coupling constants for the top-quark and the Majorana fermions run by obeying the beta functions
\al{
&(4 \pi)^{2} \beta_{y_t}= y_{t}\Bigg( \frac{9}{2} y_{t}^{2} -8 g_{S}^{2} -\frac{9}{4} g_2^{2} -\frac{17}{12} g_{Y}^{2}
-\frac{17}{12} {g}_\text{mix}^{2} \Bigg)\,, \\[2ex]
&(4 \pi)^{2} \beta_{y_R}=y_R \left( 6 y_R^{2} + 2y_L^{2}-6 g_{X}^{2}\right)\,.
}
The beta function for the left-handed Majorana Yukawa coupling constant are obtained by replacing $R\leftrightarrow L$.

For the scalar interactions, one has
\al{
(4 \pi)^{2} \beta_{\lambda_{H}}=& 24 \lambda_{H}^{2}-6 y_{t}^{4}+\frac{9}{8} g_2^{4}+\frac{3}{8} g_{Y}^{4}+\frac{3}{4} g_2^{2} g_{Y}^{2}
+\lambda_{H S}^{2}
+\frac{3}{4}{g}_\text{mix}^{2}\left(g_2^{2}+g_{Y}^{2}\right)
+\frac{3}{8}{g}_\text{mix}^{4}
+\lambda_{H}\left(12 y_{t}^{2}-9 g_2^{2}-3 g_{Y}^{2}-3{g}_\text{mix}^{2}\right) \,,\nn[2ex]
(4 \pi)^{2} \beta_{\lambda_S}=&20 \lambda_{S}^{2}-16 \left(y_{R}^{4}+y_{L}^{4}\right)+96 g_{X}^{4}
+8 \lambda_{S} \left(y_{R}^{2}+y_{L}^{2}\right)
-48 \lambda_{S} g_{X}^{2}+2 \lambda_{H S}^{2}\,,\nn[2ex]
(4 \pi)^{2} \beta_{\lambda_{H S}}=& \lambda_{H S}\bigg(12 \lambda_{H}+8 \lambda_{S}+4 \lambda_{H S}+6 y_{t}^{2}-\frac{9}{2} g_2^{2}
-\frac{3}{2} g_{Y}^{2}-\frac{3}{2}g_\text{mix}^{2}+4 \left(y_{R}^{2}+y_{L}^{2}\right)-24 g_{X}^{2} \bigg)
+12g_\text{mix}^{2} g_{X}^{2}\,.
\label{App: beta functions for scalar couplings}
}

The anomalous dimensions for the scalar masses $m_H^2$ and $m_S^2$ are given by
\al{
(4 \pi)^{2} \gamma_{m_{H}^{2}}&=m_{H}^{2}\left(12 \lambda_{H}+6 y_{t}^{2}-\frac{3}{2} g_{Y}^{2}-\frac{3}{2}g_\text{mix}^{2}\right)
+2 \lambda_{H S} m_{S}^{2}\,,\nn[2ex]
(4 \pi)^{2} \gamma_{m_{S}^{2}}&=m_{S}^{2}\left(8 \lambda_{S}+4 (y_{R}^{2}+y_{L}^{2})-24 g_{X}^{2}\right)
+4 \lambda_{H S} m_{H}^{2}\,.
\label{anomalous dimension for scalar masses}
}

\subsection{Boundary condition for gauge coupling constants}
\label{BC for gauge couplings}
We give the boundary condition for the SM gauge coupling constants.
The value of the strong coupling constant $g_S$ is extracted from $\alpha_S\fn{M_Z}=g_S^2\fn{M_Z}/4\pi=0.1184$~\cite{Tanabashi:2018oca}, where the $Z$-boson mass is $M_Z=91.1876$\,GeV.
One can obtain the values of SU(2)$_L$ and U(1)$_Y$ gauge coupling constants at $M_Z$ from the fine structure constant and the Weinberg angle~\cite{Tanabashi:2018oca} which are observed as
\al{
&\alpha\fn{M_Z}=\frac{1}{4\pi}\frac{g_Y^2\fn{M_Z}g_2^2\fn{M_Z}}{g_Y^2\fn{M_Z}+g_2^2\fn{M_Z}}=127.916\,,\nn[2ex]
&\sin^2\theta_W\fn{M_Z}=\frac{g_Y^2\fn{M_Z}}{g_Y^2\fn{M_Z}+g_2^2\fn{M_Z}}=0.23116\,.
}
From these values, more explicitly, one can extract
\al{
&g_S\fn{M_Z}=1.22029\,,&
&g_2\fn{M_Z}=0.65191\,,&
&g_Y\fn{M_Z}=0.35746\,.&
}

\section{One-loop effective potential}
\label{App: One-loop effective potential}
In this section, we give the effective potential at the one-loop level.
The Higgs and the singlet-scalar fields are parametrized by $H=(\varphi^+,h+i\varphi^0)/\sqrt{2}$ and $S=(\phi+i\eta)/\sqrt{2}$, respectively.
The effective potential at the one-loop level is
\al{
V_\text{eff}\fn{h,\phi}=V_\text{tree}\fn{h,\phi}+\Delta V_{\text{1-loop}}(h,\phi),
}
where the tree level potential is
\al{
V_\text{tree}\fn{h,\phi}=\frac{\lambda_H}{4}h^4+\frac{\lambda_{HS}}{4}h^2\phi^2+\frac{\lambda_S}{4}\phi^4\,,
}
and one has the one-loop effective potential,
\al{
\Delta V_{\text{1-loop}}(h,\phi)=& \frac{1}{64 \pi^{2}}\Bigg\{3 G_{h}^{2}\left[\ln \frac{G_{h}}{M^{2}}-\frac{3}{2}\right]+G_{\phi}^{2}\left[\ln \frac{G_{\phi}}{M^{2}}-\frac{3}{2}\right]+\operatorname{Tr}\left(H^{2}\left[\ln \frac{H}{M^{2}}-\frac{3}{2}\right]\right)\nn
&-12 T^{2}\left[\ln \frac{T}{M^{2}}-\frac{3}{2}\right]+3 \operatorname{Tr}\left(M_{G}^{2}\left[\ln \frac{M_{G}}{M^{2}}-\frac{5}{6}\right]\right)-2 \sum_{i=L,R} N_{i}^{2}\left[\ln \frac{N_{i}}{M^{2}}-\frac{3}{2}\right] \Bigg\} 
}
where $M$ is a renormalization scale, and the mass functions are defined  by 
\al{
&G_{h}(h, \phi)=\lambda_{H} h^{2}+\frac{\lambda_{HS}}{2} \phi^{2}\,,\quad
G_{\phi}(h, \phi)=\lambda_{S} \phi^{2}+\frac{\lambda_{HS}}{2} h^{2}\,,\quad
T(h, \phi)=\frac{1}{2}\left(y_{t} h\right)^{2}\,,\quad
N_{L,R}(h,\phi)=\frac{1}{2}(y_{L,R} \phi)^2\,,
\nn[2ex]
&H(h,\phi)=\pmat{
3 \lambda_{H} h^{2}+\displaystyle\frac{\lambda_{HS}}{2} \phi^{2} & {\lambda_{HS} h\, \phi} \\[2ex]
{\lambda_{HS} h\, \phi} & 3 \lambda_{S} \phi^{2}+\displaystyle\frac{\lambda_{HS}}{2} h^{2}
}\,,\quad
M_{G}(h,\phi)=\frac{1}{4}\pmat{
g_Y^{2} h^{2} & -g_2 g_Y h^{2} & g_Y g_\text{mix} h^{2} \\[2ex]
 -g_2 g_Y h^{2} & g_2^{2} h^{2} & -g_2  g_\text{mix} h^{2} \\[2ex]
g_Y  g_\text{mix} h^{2} & -g_2  g_\text{mix} h^{2} &  g_\text{mix}^{2} h^{2}+16  g_X^{2}\phi^{2}
}\,.\nn[2ex]
}
For the Higgs portal coupling and the kinetic mixing coupling to be small, one can obtain the one-loop correction from SM particles to the Higgs mass by computing
\al{
\Delta M_H^2 \simeq -\frac{\df^2 \Delta V_{\text{1-loop}}}{\df h^2}\bigg|_{h=v_h}\,.
}
\section{Cross sections for dark matter annihilation}
\label{App: Cross sections for dark matter annihilation}
We give explicit forms of thermal averaged cross sections for dark matter annihilation processes as shown in Fig.\,\ref{dark matter annihilation processes}.
To this end, we briefly summarize formulas to calculate them. 
In this section, we omit primes which denote the mass eigenstates of the scalar fields $h$ and $\phi$.

\subsection{Basic formula}
For two-body scattering process ($A+B\to a+b$) in a center-of-mass system the differential scattering cross section is given by
\al{
\label{App: two-body scattering cross section 1}
\frac{\df \sigma}{\df \Omega}\fn{A+B\to a+b}= \frac{1}{64\pi^2 s}\frac{|\bvec k|}{|\bvec p|}|\mathcal M|^2\,.
}
Here external momenta for the initial and the final states are expressed as, respectively,
\al{
|\bvec p|&=\frac{\sqrt{s-(m_A+m_B)^2}\sqrt{s-(m_A-m_B)^2}}{2\sqrt{s}}\,,&
|\bvec k|&=\frac{\sqrt{s-(m_a+m_b)^2}\sqrt{s-(m_a-m_b)^2}}{2\sqrt{s}}\,,&
}
with $s=(p_A+p_B)^2=(p_a+p_b)^2$.
Here, we assume the center-of-mass system ($p_A=(E,\bvec p)$ and $p_B=(E,-\bvec p)$) and $m_A=m_B\equiv m_i$.
With the relative velocity $v=4\sqrt{|\bvec p|^2/s}$, the cross section for the two-body scattering process  \eqref{App: two-body scattering cross section 1} is
\al{
\sigma\fn{A+B\to a+b}v= \int \df \Omega\frac{1}{16\pi^2 s}\frac{|\bvec k|}{\sqrt{s}}|\mathcal M|^2\,.
}

The thermal averaged cross section is defined by
\al{
\label{App: thermal averaged two-body scattering cross section}
\langle \sigma v\rangle= \frac{e^{2m_i /T}}{(2\pi m_i T)^3}\int \df^3 p_A \df^3 p_B \,(\sigma v)\,e^{-(E_A+E_B)/T}\,,
}
with $E_A=E_B=\sqrt{{\bvec p}^2+m_i^2}$ the energy dispersion.
The momentum integrals \eqref{App: thermal averaged two-body scattering cross section}, however, cannot by evaluated analytically, so that, by assuming a small relative velocity, one expand the cross section into a polynomial of $v^2$, i.e.,
\al{
\sigma v= \sigma^{(s)} + \sigma^{(p)} v^2+{\mathcal O}(v^{4})\,,
}
where odd power terms of $v$ are dropped since they vanish in the integrals \eqref{App: thermal averaged two-body scattering cross section}.
We obtain the formula for the thermal averaged cross section,
\al{
\label{App: thermal averaged cross section}
\langle \sigma v\rangle= \sigma^{(s)} +6\sigma^{(p)} \frac{T}{m_i} +{\mathcal O}\fn{(T/m_i)^{-2}}\,.
}
\subsection{Cross section for dark matter annihilation}
\begin{figure*}
\includegraphics[width=18.4cm]{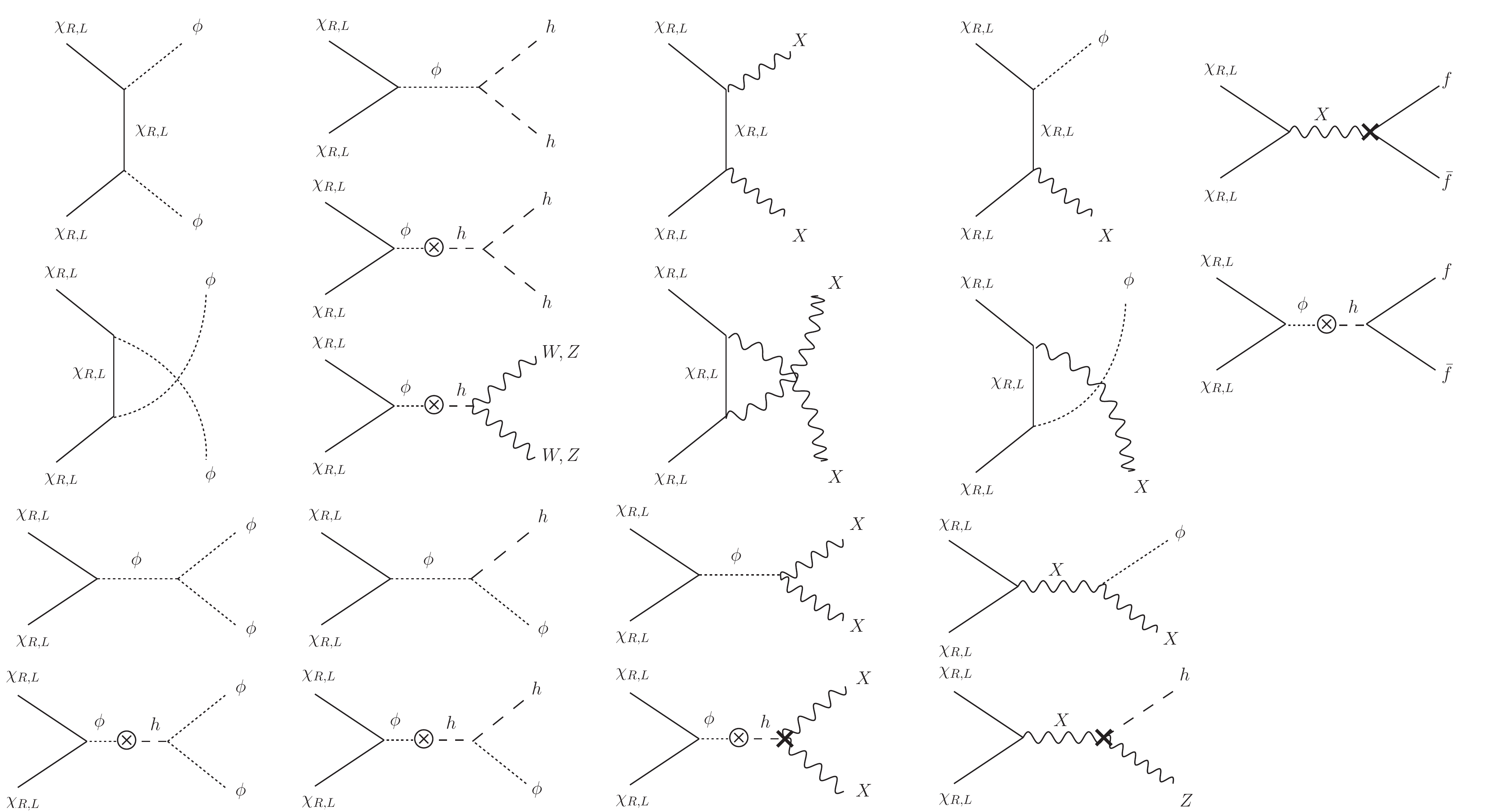}
\caption{
Possible annihilation processes of dark matters, which is denoted by $\langle \sigma\fn{\chi_L\chi_L;\text{SMs, $\phi$, $X_\mu$}} v\rangle$.
These processes are represented in the flavor basis.
The cross on the right-hand side diagram stands for the kinetic mixing between $U(1)_Y$ and $U(1)_X$ gauge fields.
The cross in a circle stands for the mixing between $\phi$ and $h$.
``SM" indicates contributions from the SM particles except for the Higgs boson, i.e., $W$ and $Z$ bosons and top-quarks, and $f$ are SM fermions.
Contributions from other quarks are neglected since their Yukawa coupling constant is relatively smaller than that of top quark, whereas for the decay process into fermions mediated by $X$ boson, one has to take into account all quark contributions.
}
\label{dark matter annihilation processes} 
\end{figure*}

Let us evaluate cross sections for each process exhibited in Fig.\,\ref{dark matter annihilation processes}.
Assuming that dark matters are non-relativistic, we expand the cross sections into a polynomial of the relative velocity $v$ and take into account up to of order $v^2$.
\begin{figure}
\includegraphics[width=9cm]{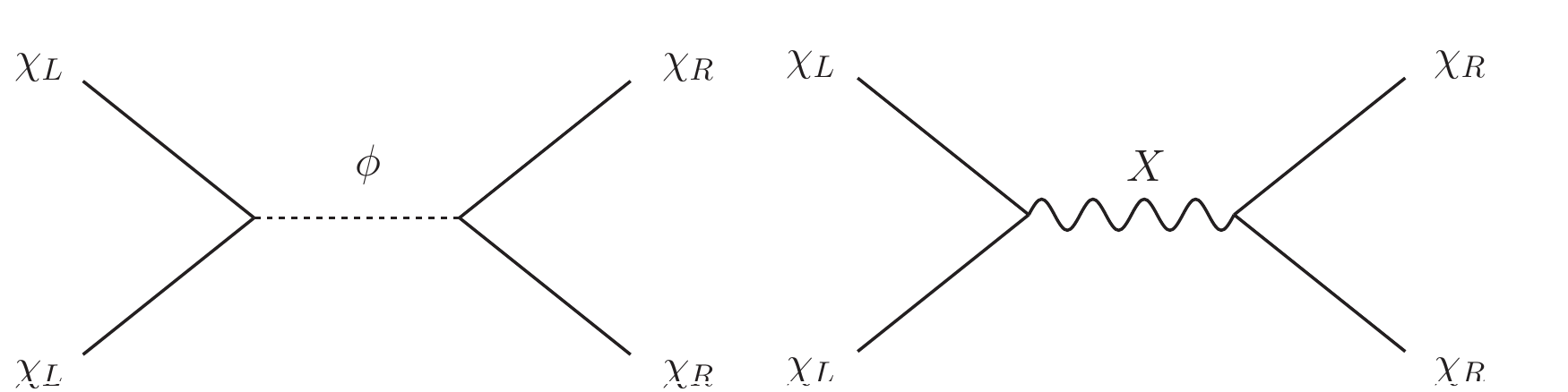}
\caption{
Annihilation processes of the light-handed Majorana fermions into the right-handed ones.
This is denoted by $\langle \sigma\fn{\chi_L\chi_L;\chi_R\chi_R} v\rangle$.
}
\label{dark matter annihilation processes LL->RR} 
\end{figure}

When two $\chi_i$s ($i=L,R$) annihilate into scalar fields, one finds
\al{
\sigma \fn{\chi_i\chi_i;\phi\phi}v&=\sigma_{\chi_i\chi_i;\phi\phi}^{(p)} v^2 + {\mathcal O}(v^4)\,,\nn[1ex]
\sigma \fn{\chi_i\chi_i;hh}v&=\sigma_{\chi_i\chi_i;hh}^{(p)} v^2 + {\mathcal O}(v^4)\,,\nn[1ex]
\sigma \fn{\chi_i\chi_i;h\phi}v&=\sigma_{\chi_i\chi_i;h\phi}^{(p)} v^2 + {\mathcal O}(v^4)\,,
}
with
\al{
&\sigma_{\chi_i\chi_i;\phi\phi}^{(p)}
=\frac{y_i^2}{16\pi}r_{i\phi}
 \left| 
 \lambda_{\text{eff},\,S^3}\,\Delta_{SS}\fn{M_i} +\frac{\lambda_{HS}v_H}{2}\Delta_{HS}\fn{M_i}\right|^2 
 +\frac{y_i^4}{6\pi}  \left|\widehat\Delta_{\phi}\fn{M_i} \right|^4 r_{i\phi}  M_i^2\left(9M_i^4-8M_i^2M_\phi^2+2M_\phi^4\right)\nn
 &\phantom{\sigma_{\chi_i\chi_i;\phi\phi}^{(p)}=} +\frac{y_i^3}{6\sqrt{2}\pi} \left|\left(3\lambda_S v_S\Delta_{SS}\fn{M_i} + \frac{\lambda_{HS}v_H}{2}\Delta_{HS}\fn{M_i}  \right)\left(\widehat\Delta_{\phi}\fn{M_i}\right)^2\right| r_{i\phi}
 M_i\left(5M_i^2-2M_\phi^2\right)\,,\\[2ex]
&\sigma_{\chi_i\chi_i;hh}^{(p)}
=\frac{y_i^2}{16\pi}r_{ih}\left| 3\lambda_H v_H\Delta_{HS}\fn{M_i}+\frac{\lambda_{HS}v_S}{2}\Delta_{SS}\fn{M_i}\right|^2\,,\\[2ex]
&\sigma_{\chi_i\chi_i;h\phi}^{(p)}
=\frac{2y_i^2}{16\pi}r_{ih\phi}
\left| \frac{\lambda_{HS}v_S}{2} \Delta_{HS}\fn{M_i}+\frac{\lambda_{HS}v_H}{2}\Delta_{SS}\fn{M_i}\right|^2\,,
}
where $\lambda_{\text{eff}S^3}$ is the effective cubic coupling constant given in Eq.\,\eqref{Eq: cubic coupling}.
Here, we define dimensionless functions,
\al{
r_{ij}&=\sqrt{1-\frac{M_j^2}{M_i^2}}\,,&
r_{ijk}&=\sqrt{1-\frac{(M_j+M_k)^2}{4M_i^2}}\sqrt{1-\frac{(M_j-M_k)^2}{4M_i^2}}\,.&
}
As given in Eq.\,\eqref{Eq: mixed propagators of scalar fields}, the propagators in the $s$ channel for the $h$-$\phi$ mixing and the singlet-scalar field $\phi$ are defined by
\al{
&\Delta_{HS}\fn{M}=\frac{\cos\theta \sin\theta}{4M^2-M_\phi^2}-\frac{\cos\theta\sin\theta}{4M^2-M_h^2+i\Gamma_H M_h}\,,
\label{propagatorhs}
\\[1ex]
&\Delta_{SS}\fn{M}=\frac{\cos^2\theta}{4M^2-M_\phi^2}-\frac{\sin^2\theta}{4M^2-M_h^2+i\Gamma_H M_h}\,,
\label{propagatorss}
}
with $\theta$ the mixing angle given in Eq.\,\eqref{mixing angle}, and $\Gamma_H$ the decay width of $H$ presented in Eq.\,\eqref{eq: decay width of Higgs}, while the $s$-channel propagator of the $X_\mu$ gauge field is given by
\al{
&\Delta_{X}\fn{M}= \frac{1}{4M^2-M_X^2}\,.
}
For annihilation processes mediated by a Majorana fermion, we define its propagators in the $t$ and $u$ channels,
\al{
&\widehat\Delta_{j}\fn{M}=\frac{1}{M_j^2-2M^2}\,,&
&\widetilde\Delta_{ij}\fn{M}=\frac{1}{M_i^2+M_j^2-4M^2}\,.&
}
Note that since the top-quark is much heavier (or equivalently larger Yukawa coupling constant) than other fermions in the SM, we neglect those of the annihilation process with Yukawa couplings of the SM.

The cross sections for the $\chi_L$ pair annihilation into SM gauge bosons, $W^+W^-$ and $ZZ$, are given by
\al{
&\sigma \fn{\chi_i\chi_i;ZZ}v=\sigma_{\chi_i\chi_i;ZZ}^{(p)} v^2 + {\mathcal O}(v^4)\,,\nn[1ex]
&\sigma \fn{\chi_i\chi_i;WW}v=\sigma_{\chi_i\chi_i;WW}^{(p)}v^2 + {\mathcal O}(v^4)\,,
}
where
\al{
&\sigma_{\chi_i\chi_i;ZZ}^{(p)} 
=\frac{y_i^2}{4\pi}
\left| \frac{M_Z^2}{v_H}\Delta_{HS}\fn{M_i}\right|^2 r_{iZ} \left( \frac{3}{4} - \frac{M_i^2}{M_Z^2} + \frac{M_i^4}{M_Z^4}\right) \,,\\[1ex]
&\sigma_{\chi_i\chi_i;WW}^{(p)} 
=\frac{y_i^2}{2\pi} \left| \frac{M_W^2}{v_H}\Delta_{HS}\fn{M_i}\right|^2 r_{iW} \left( \frac{3}{4} - \frac{M_i^2}{M_W^2} + \frac{M_i^4}{M_W^4}\right) \,,
}
while for the annihilations into SM fermions, and $Z$-Higgs pairs, one has
\al{
&\sigma \fn{\chi_i\chi_i;ff}v=\sigma_{\chi_i\chi_i;ff}^{(p)}v^2+ {\mathcal O}(v^4)\,,\nn[1ex]
&\sigma \fn{\chi_i\chi_i;h Z}v=\sigma_{\chi_i\chi_i;h Z}^{(s)}+\sigma_{\chi_i\chi_i;h Z}^{(p)}v^2+ {\mathcal O}(v^4)\,,
}
with the $s$-wave cross section for $\chi_i\chi_i\to hZ$,
\al{
&\sigma_{\chi_i\chi_i;h Z}^{(s)}=\frac{g_X^2g_\text{mix}^2}{16\pi M_X^4}r_{ihZ}^3 M_i^2\,,
}
and the $p$-wave ones,
\al{
&\sigma_{\chi_i\chi_i;ff}^{(p)}=\frac{y_i^2}{4\pi}n_{c,f}\left| \frac{M_f}{v_H}\Delta_{HS}\fn{M_i} \right|^2 r_{if} \left(M_i^2 -M_f^2\right) 
+\frac{n_{c,f} ( Y_{f_L}^2 +Y_{f_R}^2) g_\text{mix}^2 g_X^2}{24\pi}\left|\Delta_X\fn{M_i} \right|^2 r_{if}(M_f^2+2M_i^2)
\,,\\[2ex]
&\sigma_{\chi_i\chi_i;h Z}^{(p)}=\frac{g_X^2g_\text{mix}^2}{3072\pi M_i^2 M_X^4}\left|\Delta_X\fn{M_i}\right|^2 r_{ihZ} \Big[
-16 M_i^4 \Big\{18 M_h^2 \left(M_X^2-M_Z^2\right)+9 M_h^4+18 M_X^2 M_Z^2-2 M_X^4+9 M_Z^4\Big\}\nn[1ex]
&\phantom{\sigma}
+4 M_i^2 M_X^2 \Big\{M_h^2 \left(5 M_X^2-36 M_Z^2\right)+18 M_h^4+29 M_X^2
   M_Z^2+18 M_Z^4\Big\}+576 M_i^6 \left(M_h^2+M_Z^2\right)-7 M_X^4 \left(M_h^2-M_Z^2\right)^2\Big]\,.
}
Here,  $f$ denotes all fermions in the SM, $n_{c,f}$ is the degree of freedom of color, i.e. $n_{c,q}=3$ for quarks and $n_{c,\ell}=1$ for leptons, and $Y_f$ is U(1)$_Y$ hypercharge for a fermion $f$, especially, $Y_{f_L}$ and $Y_{f_R}$ stand for hypercharges of the left- and right-handed fermion sectors, respectively.

A two $\chi_L$ pair decays into the U(1)$_X$ gauge bosons whose cross section is computed as
\al{
\sigma \fn{\chi_i\chi_i;X X}v=\sigma_{\chi_i\chi_i;X X}^{(s)}+\sigma_{\chi_i\chi_i;X X}^{(p)}v^2+  {\mathcal O}(v^4)\,,
}
with the $s$- and $p$-wave cross sections,
\al{
&\sigma_{\chi_i\chi_i;X X}^{(s)}=\frac{g_X^4}{4\pi} \left|\widehat\Delta_X\fn{M_i}\right|^2r_{iX} \left( M_i^2-M_X^2\right)\,,\\[2ex]
&\sigma_{\chi_i\chi_i;X X}^{(p)}=\frac{y_i^2}{4\pi} \left| \frac{M_X^2}{v_S}\Delta_{SS}\fn{M_i}+ g_\text{mix}v_H\Delta_{HS}\fn{M_i}\right|^2 r_{iX}\left( \frac{3}{4} - \frac{M_i^2}{M_X^2} + \frac{M_i^4}{M_X^4}\right)\nn
&\qquad
+\frac{y_i}{6\sqrt{2}\pi M_X^2}\left| \left(\widehat\Delta_X\fn{M_i}\right)^2\left( \frac{M_X^2}{v_S}\Delta_{SS}\fn{M_i}+ g_\text{mix}v_H\Delta_{HS}\fn{M_i}\right)\right| r_{iX} \Bigg[ \frac{M_i}{M_X^2}\left(2M_X^6-9M_X^4M_i^2+12M_X^2M_i^4-8M_i^6\right)\Bigg]
\nn 
&\qquad
+\frac{g_X^4}{96\pi M_X^4} \left|\widehat\Delta_X\fn{M_i}\right|^4r_{iX} \Big[128M_i^{10} +17M_X^{10}-88M_X^8 M_i^2+124M_X^6 M_i^4+56M_X^4M_i^6-192M_X^2M_i^8\Big]
 \,,
}
while for the annihilation into a $X_\mu$-$\phi$ pair, one obtains
\al{
\sigma \fn{\chi_i\chi_i; X\phi}v= \sigma_{\chi_i\chi_i; X\phi}^{(s)}+\sigma_{\chi_i\chi_i; X\phi}^{(p)}v^2+{\mathcal O}(v^4)\,,
}
where
\al{
&\sigma_{\chi_i\chi_i; X\phi}^{(s)}=\frac{g_X^4}{\pi M_X^4}r_{iX\phi}^3 M_i^2\,,\\[2ex]
&\sigma_{\chi_i\chi_i; X\phi}^{(p)}=\frac{g_X^4}{192\pi M_i^2 M_X^4}\left|\Delta_X\fn{M_i}\right|^2 r_{iX\phi} \Big[
576 M_i^6 \left(M_X^2+M_{\phi }^2\right)-16 M_i^4 \left(25 M_X^4+9 M_{\phi }^4\right)\nn
&\phantom{\sigma_{\chi_i\chi_i; X\phi}^{(p)}=\frac{g_X^4}{192\pi M_i^2 M_X^4}}
+4 M_i^2 \left(-31 M_X^4 M_{\phi }^2+18 M_X^2 M_{\phi }^4+47 M_X^6\right)-7 M_X^4
   \left(M_X^2-M_{\phi }^2\right)^2\Big]\nn
&\phantom{\sigma_{\chi_i\chi_i; X\phi}^{(p)}=}
+\frac{y_i^2g_X^2}{192\pi M_i^2 M_X^2} \left|\widetilde\Delta_{X\phi}\fn{M_i}\right|^4  r_{iX\phi}\Big[
1024 M_i^{10}+256 M_i^8 \left(5 M_X^2-4 M_{\phi }^2\right)-128 M_i^6 \left(4 M_X^2 M_{\phi }^2+5 M_X^4-3 M_{\phi }^4\right)\nn
&\phantom{\sigma_{\chi_i\chi_i; X\phi}^{(p)}=\frac{g_X^4}{192\pi M_i^2 M_X^4}}
+32 M_i^4 \left(8 M_X^4 M_{\phi }^2+M_X^2 M_{\phi }^4+M_X^6-2
   M_{\phi }^6\right)+4 M_i^2 \left(M_X^4-M_{\phi }^4\right){}^2+M_X^2 \left(M_X^2-M_{\phi }^2\right)^4\Big]\nn
&\phantom{\sigma_{\chi_i\chi_i; X\phi}^{(p)}=}
+\frac{y_ig_X^3}{6\sqrt{2}\pi M_iM_X^3}\left|\Delta_X\fn{M_i}\left(\widetilde\Delta_{X\phi}\fn{M_i} \right)^2 \right|
r_{iX\phi}\Big[
128 M_i^8 - 96 M_i^6 \left(M_X^2+M_{\phi }^2\right)+8 M_i^4 \left(2 M_X^2 M_{\phi }^2+13 M_X^4+3 M_{\phi }^4\right) \nn[1ex]
&\phantom{\sigma_{\chi_i\chi_i; X\phi}^{(p)}=\frac{g_X^4}{192\pi M_i^2 M_X^4}}
- 2 M_i^2 \left(7 M_X^4 M_{\phi }^2-M_X^2 M_{\phi }^4+9M_X^6+M_{\phi }^6\right)-M_X^4 \left(M_X^2-M_{\phi }^2\right)^2   \Big]\,.
}

Finally, we show the cross section for the annihilation of $\chi_L$s into $\chi_R$s, whose process is exhibited in Fig.\,\ref{dark matter annihilation processes LL->RR}.
This results in
\al{
&\sigma\fn{\chi_L\chi_L;\chi_R\chi_R} v= \sigma_{\chi_L\chi_L;\chi_R\chi_R}^{(s)}+\sigma_{\chi_L\chi_L;\chi_R\chi_R}^{(p)}v^2 + {\mathcal O}(v^4)\,,
}
with
\al{
&\sigma_{\chi_L\chi_L;\chi_R\chi_R}^{(s)}= \frac{g_X^4}{4\pi}r_{LR}\frac{M_R^2}{M_X^4}\,, \\[2ex]
&\sigma_{\chi_L\chi_L;\chi_R\chi_R}^{(p)}=
\frac{y_L^2y_R^2}{4\pi} \left|\Delta_{SS}\fn{M_L}\right|^2r_{LR} \left(M_L^2-M_R^2\right)
+\frac{g_X^4}{96\pi M_X^4\left(M_L^2-M_R^2 \right)}\left|\Delta_{X}\fn{M_L} \right|^2r_{LR}\Big\{
-48 M_L^4 M_R^2 M_X^2\nn
&\phantom{\sigma_{\chi_L\chi_L;\chi_R\chi_R}^{(p)}=\frac{3y_L^2y_R^2}{192\pi}}
+22 M_L^2 M_R^2 M_X^4+72 M_L^2 M_R^4 M_X^2+96 M_L^6 M_R^2-144 M_L^4 M_R^4-8 M_L^4 M_X^4-17 M_R^4 M_X^4\Big\}
 \,.
}

\subsection{Thermal average cross section}
Utilizing the formula \eqref{App: thermal averaged cross section} the thermal averaged cross sections for dark matter annihilation to SM particles, $\phi$ and $X_\mu$ are given by
\al{
\langle \sigma\fn{\chi_i\chi_i;\text{SMs, $h$, $\phi$}} v\rangle&=
\frac{1}{16\pi}\Bigg[ 
r_{iX} a_{X} \fn{g_X,M_i}
+r_{ihZ}\, a_{hZ}\fn{g_X,M_i}
+r_{iX\phi}\, a_{X\phi}\fn{g_X,M_i}\Bigg]\nn
&\quad
+
\frac{1}{16 \pi }\frac{3\mu_{RL}}{xM_i} \Bigg[ 
 \sum_{I=W, Z, t, h, \phi}r_{iI}\, b_{I}\left(y_i, M_{i}\right) +\sum_{f}r_{if}b_{f}\fn{y_i,g_X,M_i}  +r_{iX}\,b_X\fn{y_i,g_X,M_i}  \nn
&\phantom{\langle \sigma\fn{\chi_i\chi_i;\text{SMs, $h$, $\phi$}} v\rangle}
 +r_{ih\phi}\, b_{h\phi}\fn{y_i,M_i}  +r_{ihZ} b_{hZ}\fn{g_X,M}+ r_{iX\phi} b_{X\phi}\fn{y_i,g_X,M} \Bigg]\,,
}
with $1/\mu_{RL}=1/M_R+1/M_L$.
Here, the five diagrams of the right-hand side in Fig.\,\ref{dark matter annihilation processes} give contributions $b_I$ so that
\al{
&a_X\fn{\kappa,M}= 4g_X^2\left(\kappa\right)^2 \left|\widehat\Delta_X\fn{M}\right|^2\left( M^2-M_X^2\right)\,\nn[2ex]
&a_{hZ}\fn{\kappa,M}=\frac{g_\text{mix}^2}{M_X^4} \left(\kappa\right)^2 \left( 1-\frac{(M_h+M_Z)^2}{4M^2}\right)\left( 1-\frac{(M_h-M_Z)^2}{4M^2} \right) M^2\,,\nn[2ex]
&a_{X\phi}\fn{\kappa,M}=\frac{16g_X^2}{M_X^4}\left(\kappa\right)^2  \left( 1-\frac{(M_X+M_\phi)^2}{4M^2}\right)\left( 1-\frac{(M_X-M_\phi)^2}{4M^2} \right) M^2\,.
}
The coefficients $a_I$ correspond to the diagrams the three left lines of Fig.\,\ref{dark matter annihilation processes}.
They result in
\al{
&b_W\fn{\kappa,M}=16\left(\kappa \right)^2 \left| \frac{ M_W^2}{v_H}\Delta_{HS}\fn{M} \right|^2\left( \frac{3}{4}-\frac{M^2}{M_W^2}+\frac{M^4}{M_W^4}\right),\nn[2ex]
&b_Z\fn{\kappa,M}=8\left(\kappa\right)^2 \left| \frac{ M_Z^2}{v_H}\Delta_{HS}\fn{M} \right|^2\left( \frac{3}{4}-\frac{M^2}{M_Z^2} +\frac{M^4}{M_Z^4} \right),\nn[2ex]
&b_f\fn{\kappa,\rho,M}=8n_{c,f}\left(\kappa \right)^2 \left|\frac{ M_f}{v_H}\Delta_{HS}\fn{M} \right|^2  \left(M^2-M_f^2\right)
+\frac{16n_{c,f}(Y_{f_L}^2+Y_{f_R}^2) g_\text{mix}^2}{3} (\rho)^2\left|\Delta_X\fn{M} \right|^2 \left( M_f^2 +2M^2 \right) \,,\nn[2ex]
&b_h\fn{\kappa,M}=2\left( \kappa\right)^2 \left| 3\lambda_H v_H\Delta_{HS}\fn{M}+\frac{\lambda_{HS}v_S}{2}\Delta_{SS}\fn{M}\right|^2\,,\nn[2ex]
&b_\phi\fn{\kappa,M}=2\left( \kappa\right)^2 \left| 3\lambda_S v_S\Delta_{SS}\fn{M} +\frac{\lambda_{HS}v_H}{2}\Delta_{HS}\fn{M} \right|^2
 +\frac{16}{3}(\kappa)^4 \left|\widehat\Delta_{\phi}\fn{M} \right|^4  M^2\left(9M^4-8M^2M_\phi^2+2M_\phi^4\right)\nn
 &\phantom{a_\phi\fn{\kappa,M}} 
 +\frac{16}{3\sqrt{2}}(\kappa)^3 \left|\left(3\lambda_S v_S\Delta_{SS}\fn{M} + \frac{\lambda_{HS}v_H}{2}\Delta_{HS}\fn{M}  \right)\left(\widehat\Delta_{\phi}\fn{M}\right)^2\right| M\left(5M^2-2M_\phi^2\right)
\,,\nn[2ex]
&b_{h\phi}\fn{\kappa,M}=4\left( \kappa\right)^2\left| \frac{\lambda_{HS}v_S}{2} \Delta_{HS}\fn{M_L}+\frac{\lambda_{HS}v_H}{2}\Delta_{SS}\fn{M_L}\right|^2\,,\nn[2ex]
&b_X\fn{\kappa,\rho,M}=8\left(\kappa\right)^2\left|
 \frac{M_X^2}{v_S} \Delta_{SS}\fn{M} +  g_\text{mix}v_H\Delta_{HS}\fn{M}
\right|^2 \left( \frac{3}{4}-\frac{M^2}{M_X^2} +\frac{M^4}{M_X^4} \right)\nn
&\phantom{a_X{\kappa,\rho,M}}
+\frac{16}{3 \sqrt{2}M_X^2}(\kappa) \left| \left(\widehat\Delta_X\fn{M}\right)^2\left( \frac{M_X^2}{v_S}\Delta_{SS}\fn{M}+ g_\text{mix}v_H\Delta_{HS}\fn{M}\right)\right|  \Bigg[ \frac{M}{M_X^2}\left(2M_X^6-9M_X^4M^2+12M^4M_X^2-8M^6\right)\Bigg]
\nn 
&\phantom{a_X{\kappa,\rho,M}}
+\frac{1}{3 M_X^4}(\rho)^4 \left|\widehat\Delta_X\fn{M}\right|^4\Big[128M^{10} +17M_X^{10}-88M_X^8 M^2+124M_X^6 M^4+56M_X^4M^6-192M_X^2M^8\Big]\,, \nn[2ex]
&b_{hZ}\fn{\kappa,M}=\frac{g_\text{mix}^2}{96 M^2 M_X^4}\left(\kappa\right)^2\left|\Delta_X\fn{M}\right|^2 \Big[
576 M^6 \left(M_h^2+M_Z^2\right)\nn
&\phantom{b_{hZ}\fn{\kappa,M}=\frac{g_\text{mix}^2}{3072\pi M^2 M_X^4}}
-16 M^4 \Big\{18 M_h^2 \left(M_X^2-M_Z^2\right)+9 M_h^4+18 M_X^2 M_Z^2-2 M_X^4+9 M_Z^4\Big\}\nn[1ex]
&\phantom{b_{hZ}\fn{\kappa,M}=\frac{g_\text{mix}^2}{3072\pi M^2 M_X^4}}
+4 M^2 M_X^2 \Big\{M_h^2 \left(5 M_X^2-36 M_Z^2\right)+18 M_h^4+29 M_X^2
   M_Z^2+18 M_Z^4\Big\}-7 M_X^4 \left(M_h^2-M_Z^2\right)^2\Big], \nn[2ex]
&b_{X\phi}\fn{\kappa,\rho,M}=
\frac{1}{6 M^2 M_X^4}\left(\rho\right)^4\left|\Delta_X\fn{M}\right|^2 \Big[
576 M^6 \left(M_X^2+M_{\phi }^2\right)-16 M^4 \left(25 M_X^4+9 M_{\phi }^4\right)\nn
&\phantom{\sigma_{\chi_i\chi_i; X\phi}^{(p)}=\frac{g_X^4}{192\pi M_i^2 M_X^4}}
+4 M^2 \left(-31 M_X^4 M_{\phi }^2+18 M_X^2 M_{\phi }^4+47 M_X^6\right)-7 M_X^4
   \left(M_X^2-M_{\phi }^2\right)^2\Big]\nn
&\phantom{\sigma_{\chi_i\chi_i; X\phi}^{(p)}=}
+\frac{1}{6 M^2 M_X^2}\left(\kappa\rho\right)^2 \left|\widetilde\Delta_{X\phi}\fn{M}\right|^4 \Big[
1024 M^{10}+256 M^8 \left(5 M_X^2-4 M_{\phi }^2\right)-128 M^6 \left(4 M_X^2 M_{\phi }^2+5 M_X^4-3 M_{\phi }^4\right)\nn
&\phantom{\sigma_{\chi_i\chi_i; X\phi}^{(p)}=\frac{g_X^4}{192\pi M_i^2 M_X^4}}
+32 M^4 \left(8 M_X^4 M_{\phi }^2+M_X^2 M_{\phi }^4+M_X^6-2
   M_{\phi }^6\right)+4 M^2 \left(M_X^4-M_{\phi }^4\right){}^2+M_X^2 \left(M_X^2-M_{\phi }^2\right)^4\Big]\nn
&\phantom{\sigma_{\chi_i\chi_i; X\phi}^{(p)}=}
+\frac{16}{3\sqrt{2} MM_X^3}\left(\kappa\rho^3 \right)\left|\Delta_X\fn{M}\left(\widetilde\Delta_{X\phi}\fn{M} \right)^2 \right|
\Big[
128 M^8 - 96 M^6 \left(M_X^2+M_{\phi }^2\right)
+8 M^4 \left(2 M_X^2 M_{\phi }^2+13 M_X^4+3 M_{\phi }^4\right)\nn[1ex]
&\phantom{\sigma_{\chi_i\chi_i; X\phi}^{(p)}=\frac{g_X^4}{192\pi M_i^2 M_X^4}}
 - 2 M^2 \left(7 M_X^4 M_{\phi }^2-M_X^2 M_{\phi }^4+9M_X^6+M_{\phi }^6\right)-M_X^4 \left(M_X^2-M_{\phi }^2\right)^2   \Big].
\label{Coefficients for annihilation processes via S and H}
}

The $\chi_L\chi_L\to \chi_R\chi_R$ process is evaluated as
\al{
&\langle \sigma\fn{\chi_L\chi_L;\chi_R\chi_R} v\rangle=
\frac{g_X^4}{4\pi}r_{LR}\frac{M_R^2}{M_X^4}\nn
&+\frac{r_{LR}}{16\pi}\frac{3\mu_{RL}}{x M_{L}}\Bigg[
8(y_Ry_L)^2|\Delta_{SS}\fn{M_L}|^2  \left(M_{L}^2-M_{R}^2\right)
+\frac{g_X^4}{3M_X^4(M_L^2-M_R^2)}\left|\Delta_{X}\fn{M_L} \right|^2\Big\{
-48 M_L^4 M_R^2 M_X^2\nn
&\phantom{\sigma_{\chi_L\chi_L;\chi_R\chi_R}^{(p)}=\frac{3y_L^2y_R^2}{192\pi}}
+22 M_L^2 M_R^2 M_X^4+72 M_L^2 M_R^4 M_X^2+96 M_L^6 M_R^2-144 M_L^4 M_R^4-8 M_L^4 M_X^4-17 M_R^4 M_X^4\Big\}\Bigg]\,.
}
\twocolumngrid
\bibliography{refs}
\end{document}